\documentclass{article}
\usepackage[dvips]{graphicx} 
\usepackage{slashed,booktabs,amsthm}
\usepackage{amsmath}
\usepackage{float,bm,physics,slashed}
\usepackage{amssymb}
\usepackage[caption=false]{subfig}
\usepackage{amsmath}
\usepackage{float}
\usepackage{amssymb}
\usepackage[caption=false]{subfig}
\usepackage[colorlinks=true,linkcolor=blue,filecolor=blue,urlcolor=blue,citecolor=blue]{hyperref}
\usepackage{geometry} 
\usepackage{array,appendix}    
\usepackage{calc}
\usepackage{accents}

\usepackage[utf8]{inputenc}
\usepackage{amsmath,mathrsfs}
\usepackage{slashed,csquotes}
\usepackage{xcolor,longtable}
\usepackage{tabularx,colortbl}
\usepackage{matlab-prettifier} 
\usepackage[export]{adjustbox}
\usepackage{graphicx}
\usepackage{csquotes}
\usepackage{array}
\newcommand{\mycomment}[1]{}
\usepackage{tikz}
\usepackage{threeparttable}
\newcolumntype{P}[1]{>{\centering\arraybackslash}p{#1}}
\usepackage{mciteplus}

\begin{document}

\title{HQET Spectroscopy of Radially Excited F-wave Mesons}
\author{ Palak Gupta$^{a}$, Sapana Yadav$^{a}$, Ritu Garg$^{a}$
\\\small{\it $^{a}$Department of Physics, Manipal University Jaipur, Jaipur} \\\small{E-mail: ritu.garg@jaipur.manipal.edu}}

\maketitle

\begin{abstract}
 Motivated by the continuous advances in modern experimental facilities, we investigate the radially excited F-wave bottom and bottom-strange mesons that have not yet been observed experimentally. By combining theoretical inputs with available experimental information on charm mesons and applying flavor-symmetry parameters, we predict the masses of the radially excited F-wave bottom states and their strange partners. In addition, we compute their strong decay widths in terms of the hadronic coupling constants $\tilde{g}_{ZH}$ and $\tilde{g}_{RH}$. Using the presently estimated values of these couplings, we obtain upper bounds on the corresponding decay widths. These predictions will serve as testable benchmarks for forthcoming heavy-flavor spectroscopy findings.
\end{abstract}
\section{Introduction}
In the traditional quark model, hadrons are formed from quarks: a quark-antiquark ($ q\bar {q}$) pair forms mesons, while baryons are formed from three quarks ($qqq$). The traditional quark model can describe most hadrons, but some show exotic features that it cannot explain. This indicates that hadrons may have a more complex structure, and further study is needed to understand their inner nature. 
The existence of the bottom quark was confirmed in 1977 at Fermilab, when a resonance near $9.5\,\text{GeV}$ appeared in dimuon events from proton-nucleus collisions. This signal was subsequently identified as the $\Upsilon(1S)$ bottomonium state, corresponding to a $b\bar{b}$ bound system \cite{Herb1977}. This provided the first direct evidence for the fifth quark and a key validation of the Standard Model. In 1983, the CLEO collaboration reported ground-state bottom mesons $B^{+}$ and $B^{0}$ produced in strong decay of $Y(4S)$ \cite{Behrends1983}. Subsequently, in 1993, the first bottom-strange was independently observed by ALEPH \cite{Buskulic1993} and CDF experiment \cite{Abe1993}, respectively. Within the heavy-light meson sector, the $B$ and $B_s$ states are especially intriguing because the bottom quark mass is substantially higher than that of the charm quark. This pronounced mass hierarchy makes bottom and bottom-strange mesons an ideal sector for examining the validity of the heavy quark effective theory framework.

Despite the first bottom meson being discovered thirty years ago, the family of bottom mesons remains sparsely populated. Only seven bottom mesons are currently listed in the Review of Particle Physics, namely 
$B(5279)$, $B^*(5325)$, $B_1(5721)$, $B_J(5732)$, $B_2(5747)$, $B_J(5840)$, and $B_J(5970)$ \cite{PDG2024}. The first two bottom mesons, $B(5279)$ and $B^*(5325)$, are well established as the $1S$ states with quantum numbers $J=0$ and $J=1$, respectively. In 1994, the OPAL Collaboration observed an excess in the $B^+\pi^{-}$ invariant mass spectrum near $5.60- 5.85 \text{GeV}$, later identified as $B_J(5732)$ \cite{Abreu1995}. Several other experiments subsequently confirmed this structure and have been interpreted as arising from a combination of narrow and broad bottom meson states \cite{Buskulic1996, Barate1998, Acciarri1999, Affolder2001}. The $P$-wave bottom states $B_1(5721)$ and $B_2(5747)$ were first observed directly in 2007 by the D0 Collaboration through their decays into $B^{+(*)}\pi$ \cite{Abazov2007}. Their existence was subsequently confirmed in 2008 by the CDF Collaboration \cite{aaltonen2009}. Based on the complete Run II dataset, the CDF Collaboration in 2013 not only reaffirmed the presence of $B_1(5721)$ and $B_2(5747)$ but also reported indications of a new resonance, identified as $B(5970)$ \cite{aaltonen2014}. 
A year later, the LHCb Collaboration carried out a detailed study of the invariant mass spectra of $B^+\pi^-$ and $B^0\pi^+$, leading to high-precision determinations of the resonance parameters of $B_1(5721)$ and $B_2(5747)$ \cite{aaij2015b}. Moreover, resonant structures appearing in the 5800–6000~MeV mass region were assigned to $B_J(5840)$ and $B_J(5900)$ \cite{aaij2015b}.

Similarly to the bottom mesons, the bottom-strange meson family remains sparsely populated. According to the Review of Particle Physics, seven bottom-strange mesons have been identified: $B_s^0(5367)$, $B_s^{*0}(5415)$, $B_{s1}^0(5830)$, $B_{s2}^{*0}(5840)$, $B_{sJ}^{*0}(5850)$, $B_{sJ}^0(6063)$, and $B_{sJ}^0(6114)$ \cite{PDG2024}. The ground $S$-wave states, $B_s^0(5367)$ and $B_s^{*0}(5415)$, are well established. Analogous to $B_J(5732)$, the OPAL Collaboration reported evidence for a structure, denoted as $B_{sJ}^{*0}(5850)$, in the $B^+K^-$ invariant mass spectrum in the 5.80–6.00 $GeV$ range \cite{Akers1995}. Using $p\bar{p}$ collisions at $\sqrt{s} = 1.96$ $TeV$ collected with the CDF II detector at the Fermilab Tevatron, the CDF Collaboration observed for the first time the narrow $j_q = 3/2$ orbitally excited bottom-strange mesons: $B_{s1}^0(5830)$ in the $B^{*+}K^-$ channel and $B_{s2}^{*0}(5840)$ in the $B^+K^-$ channel \cite{aaltonen2008}. Around the same period, the D0 Collaboration also reported the direct observation of the $L=1$ excited state $B_{s2}^{*0}(5840)$ in the $B^+K^-$ invariant mass distribution with a statistical significance exceeding $4.8\sigma$, while the $B_{s1}^0(5830)$ signal was observed with a significance below $3\sigma$ \cite{abazov2008}.

In Ref.\cite{aaij2013}, the LHCb Collaboration reported the first observation of the decay $B_{s2}^{*0}(5840) \to B^{*+}K^-$. Later, the CMS Collaboration measured both $B_{s2}^{*0}(5840)$ and $B_{s1}^0(5830)$ in the $B^{(*)+}K^-$ and $B^{(*)0}K^0_s$ channels, determining the branching fraction of the neutral decay relative to the charged decay for $B_{s2}^{*0}(5840)$ \cite{sirunyan2018}. The two most recently discovered bottom-strange mesons, $B_{sJ}^0(6063)$ and $B_{sJ}^0(6114)$, were observed by the LHCb Collaboration in 2020 through the $B^\pm K^\mp$ invariant mass distributions \cite{aaij2021}. The LHCb Collaboration proposed these newly observed $B_{sJ}(6063)$ and $B_{sJ}(6114)$ states as $D$-wave excitations. These results, together with previous observations of radially excited states, highlight ongoing progress in heavy meson spectroscopy and suggest that more excited bottom mesons may be identified in future experiments.

From the theoretical perspective, numerous investigations have addressed the spectroscopy of highly excited bottom non-strange and bottom-strange mesons within a variety of frameworks \cite{13a,14a,15a,16a,17a,18a,19a,20a,21a,22a,23a,24a,25a,26a,27a,28a,29a,30a,31a,32a,33a}. In particular, the ground states $B^{0,\pm}(5279)$, $B^{*}(5324)$, $B_{s}(5366)$, and $B_{s}^{*}(5415)$ are consistently classified as $1S$ states, in full agreement with experimental determinations. Moreover, the $P$-wave states $B_{1}(5721)$ and $B_{2}^{*}(5747)$ have been firmly established, corresponding to the $1P$ doublet with quantum numbers $1^{+}$ and $2^{+}$, respectively.  

Nevertheless, the internal structure of the $B_{1}(5721)$ remains a subject of debate. Heavy meson effective theory supports its interpretation as a $1P(1^{+})$ state \cite{25a,32a}, whereas relativistic and non-relativistic quark model calculations favor a mixed configuration involving $^3P_1$ and $^1P_1$ components \cite{14a,16a,33a}. The quantum assignment of the $B_J(5840)$ also remains unsettled. Within quark potential models, it has been suggested to correspond to the $2^{1}S_{0}$ excitation \cite{16a,17a}, while analyses based on the $^3P_0$ decay model favor a $2^{3}S_{1}$ interpretation \cite{20a}. Alternatively, heavy-quark effective theory predicts its structure to be compatible with a $1^{3}D_{1}$ state \cite{28a}. A similar ambiguity exists for the $B_J(5960)^{0,+}$ resonance, which has been interpreted across different theoretical frameworks as either $2^{3}S_{1}$, $1^{3}D_{3}$, or $1^{3}D_{1}$~\cite{14a,15a,16a,22a,23a,24a,34a}. Its spin-parity quantum numbers are yet to be determined, as the PDG currently lists only its mass and decay width. Related assignments for $B_{1}(5721)$, $B_{2}^{*}(5747)$, and $B_J(5970)$ have also been discussed in our earlier work~\cite{28a}.  

In the bottom-strange sector, experimental observations remain comparatively sparse. Among the established states, $B_{s1}(5830)$ and $B_{s2}(5840)$ have been unambiguously identified by the CDF, D0, and LHCb Collaborations, and are consistently assigned as members of the $1P$ multiplet with $J^{P}=1^{+}$ and $2^{+}$, respectively~\cite{11}. However, the more recently reported resonances $B_{sJ}(6063)$ and $B_{sJ}(6114)$ have generated considerable discussion. Within the non-relativistic quark potential framework, these states are interpreted as $1^{3}D_{1}$ and $1^{3}D_{3}$ excitations, while Ref.~\cite{34a} proposes alternative assignments as $1^{3}D_{1}$ and $2^{3}S_{1}$, respectively. Theoretical understanding of these new bottom-strange states remains limited, underlining the importance of further studies in this sector.

A look through of available literature reveals that the experimental identification of higher orbital and radial excitations in the bottom-meson spectrum, including their strange counterparts, remains uncovered. The present theoretical study is therefore motivated by this scarcity of experimental information on several bottom states, particularly those with strangeness. In this work, we examine the properties of these yet-unobserved bottom mesons, specifically their masses, decay widths, and upper bounds within the framework of heavy-quark effective theory (HQET). 

This paper is structured as follows. Section~2 describes the HQET formalism adopted for the analysis of strong decay channels. Section~3 presents and discusses the numerical results. Finally, Section~4 provides the concluding remarks of this study.

\section{Framework}

The spectroscopy properties of heavy-light hadrons, including masses, decay widths, branching ratios, and quantum numbers (spin and parity), can be systematically investigated through Heavy Quark Effective Theory (HQET) framework. This theoretical approach provides a powerful tool for describing hadrons that contain a single heavy quark.

HQET incorporates two approximate symmetries: heavy quark symmetry and chiral symmetry. The heavy quark symmetry becomes exact in the limit where the heavy quark mass $m_Q \to \infty$. In this limit, the spin of the light degrees of freedom (light quarks and gluons) decouples from the spin of the heavy quark. Consequently, the total angular momentum of the light components remains conserved and can be written as
\[
s_l = s_q + l,
\]
where $s_q = \frac{1}{2}$ denotes the spin of the light quark and $l$ represents its total orbital angular momentum.

Under the heavy quark limit, mesons organize into degenerate doublets, classified according to the total angular momentum $s_l$ of the light degrees of freedom:

\begin{itemize}
    \item For $l = 0$, we have $s_l = \frac{1}{2}$. Coupling this with the heavy quark spin $s_Q = \frac{1}{2}$ yields a doublet with quantum numbers $(0^-, 1^-)$, represented by $(P, P^*)$.
    \item For $l = 1$, two doublets appear: $(P_0^*, P_1')$ with $J_{s_l}^P = (0^+, 1^+)$, and $(P_1, P_2^*)$ with $J_{s_l}^P = (1^+, 2^+)$.
    \item For $l = 2$, the corresponding doublets are $(P_1^*, P_2)$ and $(P_2', P_3^*)$, characterized by $J_{s_l}^P = (1^-, 2^-)$ and $J_{s_l}^P = (2^-, 3^-)$, respectively.
    \item For $l = 3$, the doublets $(P_2^*, P_3)$ and $(P_3', P_4^*)$ arise, associated with $J_{s_l}^P = (1^-, 2^-)$ and $J_{s_l}^P = (2^-, 3^-)$, respectively.
\end{itemize}

These doublets are conveniently represented in terms of \textbf{superfield operators}, which encapsulate the heavy--light meson states in a compact formalism. The relevant superfields are denoted as
\[
H_a, \quad S_a, \quad T_a, \quad X_a^{\mu}, \quad Y_a^{\mu\nu}, \quad Z_a^{\mu\nu}, \quad \text{and} \quad R_a^{\mu\nu\rho},
\]
each corresponding to a specific doublet structure associated with a given orbital excitation $l$. The expressions for the described fields are given below:

\begin{gather}
\label{eq:lagrangian}
 H_{a}=\frac{1+\slashed
v}{2}\{P^{*}_{a\mu}\gamma^{\mu}-P_{a}\gamma_{5}\}\\
S_{a} =\frac{1+\slashed v}{2}[{P^{'\mu}_{1a}\gamma_{\mu}\gamma_{5}}-{P_{0a}^{*}}]\\
T^{\mu}_{a}=\frac{1+\slashed v}{2}
\{P^{*\mu\nu}_{2a}\gamma_{\nu}-P_{1a\nu}\sqrt{\frac{3}{2}}\gamma_{5}
[g^{\mu\nu}-\frac{\gamma^{\nu}(\gamma^{\mu}-\upsilon^{\mu})}{3}]\}
\end{gather}
\begin{gather}
X^{\mu}_{a}=\frac{1+\slashed
v}{2}\{P^{\mu\nu}_{2a}\gamma_{5}\gamma_{\nu}-P^{*}_{1a\nu}\sqrt{\frac{3}{2}}[g^{\mu\nu}-\frac{\gamma_{\nu}(\gamma^{\mu}+v^{\mu})}{3}]\}
\end{gather}

\begin{multline}
 Y^{\mu\nu}_{a}=\frac{1+\slashed
v}{2}\{P^{*\mu\nu\sigma}_{3a}\gamma_{\sigma}-P^{'\alpha\beta}_{2a}\sqrt{\frac{5}{3}}\gamma_{5}[g^{\mu}_{\alpha}g^{\nu}_{\beta}
-\frac{g^{\nu}_{\beta}\gamma_{\alpha}(\gamma^{\mu}-v^{\mu})}{5}-\frac{g^{\mu}_{\alpha}\gamma_{\beta}(\gamma^{\nu}-v^{\nu})}{5}]\}   
\end{multline}
\begin{multline}
Z^{\mu\nu}_{a}=\frac{1+\slashed
v}{2}\{P^{\mu\nu\sigma}_{3a}\gamma_{5}\gamma_{\sigma}-P^{*\alpha\beta}_{2a}\sqrt{\frac{5}{3}}[g^{\mu}_{\alpha}g^{\nu}_{\beta}
-\frac{g^{\nu}_{\beta}\gamma_{\alpha}(\gamma^{\mu}+v^{\mu})}{5}-\frac{g^{\mu}_{\alpha}\gamma_{\beta}(\gamma^{\nu}+v^{\nu})}{5}]\}
\end{multline}
\begin{multline}
R^{\mu\nu\rho}_{a}=\frac{1+\slashed
v}{2}\{P^{*\mu\nu\rho\sigma}_{4a}\gamma_{5}\gamma_{\sigma}-P^{'\alpha\beta\tau}_{3a}\sqrt{\frac{7}{4}}[g^{\mu}_{\alpha}g^{\nu}_{\beta}g^{\rho}_{\tau}\\
-\frac{g^{\nu}_{\beta}g^{\rho}_{\tau}\gamma_{\alpha}(\gamma^{\mu}-v^{\mu})}{7}-\frac{g^{\mu}_{\alpha}g^{\rho}_{\tau}\gamma_{\beta}(\gamma^{\nu}-v^{\nu})}{7}-\frac{g^{\mu}_{\alpha}g^{\nu}_{\beta}\gamma_{\tau}(\gamma^{\rho}-v^{\rho})}{7}]\}
\end{multline}
The field $H_a$ shows doublets of S-wave for $J^P = (0^-,1^-)$. The fields $S_a$ and $T_a$ describes doublets of P-wave for $J^P = (0^+, 1^+)$ and $(1^+, 2^+)$ respectively. D-wave doublets for $J^P = (1^-, 2^-)$ and $(2^-, 3^-)$ belongs to fields $X^{\mu}_{a}$ and $Y^{\mu\nu}_{a}$ respectively. In same manner, fields $Z^{\mu\nu}_{a}$, $R^{\mu\nu\rho}_{a}$ presents doublets of F-wave for $J^P = (2^+, 3^+)$ and $(3^+, 4^+)$ respectively. $a$ in above expressions is light quark ($u, d, s$) flavor index. $v$ is heavy quark velocity, conserved in strong interactions. The approximate chiral symmetry $SU(3)_L\times SU(3)_R$ is involved with fields of pseudoscalar mesons $\pi$, K, and $\eta$ which are lightest strongly interacting  bosons. They are treated as approximate Goldstone bosons of this chiral symmetry and can be introduced by the matrix field $U(x) = Exp\left[\dot{\iota}\sqrt{2}\phi(x)/f\right]$, where $\phi(x)$ is given by
\begin{align}
\phi(x) =
\begin{pmatrix}
\frac{1}{\sqrt{2}}\pi^0+\frac{1}{\sqrt{6}}\eta & \pi^+ & K^+\\
\pi^-& -\frac{1}{\sqrt{2}}\pi^0+\frac{1}{\sqrt{6}}\eta & K^0\\
K^- & \bar{K}^0 & -\sqrt{\frac{2}{3}}\eta
\end{pmatrix}
\end{align}
The fields of heavy meson doublets (1-7) interact with pseudoscalar goldstone bosons via covariant derivative $D_{\mu ab}= -\delta_{ab}\partial_{\mu}+\mathcal{V}_{\mu ab} =  -\delta_{ab}\partial_{\mu}+\frac{1}{2}(\xi^{+}\partial_{\mu}\xi+\xi\partial_{\mu}\xi^{+})_{ab}$ and axial vector field $A_{\mu ab}=\frac{i}{2}(\xi\partial_{\mu}\xi^{\dag}-\xi^{\dag}\partial_{\mu}\xi)_{ab}$. By including all meson doublet fields and goldstone fields, effective lagranigan is given by:
\begin{multline}
    \mathcal{L} = iTr[\bar{H}_{b}v^{\mu}D_{\mu ba}H_{a}] +  \frac{f_\pi^{2}}{8}Tr[\partial^{\mu}\Sigma\partial_{\mu}\Sigma^{+}] + Tr[\bar{S_{b}}(iv^{\mu}D_{\mu ba} - \delta_{ba}\Delta_{S})S_{a}]+Tr[\bar{T_{b}^{\alpha}}(iv^{\mu}D_{\mu ba}- \delta_{ba}\Delta_{T})\\T_{a \alpha}- Tr[\bar{X_{b}^{\alpha}}(iv^{\mu}D_{\mu ba}- \delta_{ba}\Delta_{X})X_{a \alpha}+ Tr[\bar{Y_{b}^{\alpha\beta}}(iv^{\mu}D_{\mu ba}- \delta_{ba}\Delta_{Y})Y_{a\alpha\beta}+\\Tr[\bar{R_{b}^{\alpha\beta\rho}}(iv^{\mu}D_{\mu ba}- \delta_{ba}\Delta_{R})T_{a \alpha\beta\rho}] 
\end{multline}
The mass parameter $\Delta_{F}$ in equation (9) presents the mass difference between higher mass doublets (F) and lowest lying doublet (H) in terms of spin average masses of these doublets with same principle quantum number (n). The expressions for mass parameters are given by:
\begin{align}
             \Delta_{F}=\overline{M_{F}}&- \overline{M_{H}},~~ F= S,T,X,Y,Z,R\\
\text{where, }~~~~~~~~~~~
           \overline{M_{H}}&=(3m^{Q}_{P_1^*}+m^{Q}_{P_{0}})/4\\
         \overline{M_{S}}&=(3m^{Q}_{P_1^{'}}+m^{Q}_{P_0^*})/4\\
        \overline{M_{T}}&=(5m^{Q}_{P_2^*}+3m^{Q}_{P_1})/8\\
          \overline{M_{X}}&=(5m^{Q}_{P_2}+3m^{Q}_{P_1^{*}})/8\\
           \overline{M_{Y}}&=(5m^{Q}_{P_3^*}+3m^{Q}_{P_2^{'}})/8\\
            \overline{M_{Z}}&=(7m^{Q}_{P_3}+5m^{Q}_{P_2^*})/12\\
             \overline{M_{R}}&=(9m^{Q}_{P_4^*}+7m^{Q}_{P_{3}^{'}})/12
         \label{averagemassequation}
         \end{align}
           The $1/m_{Q}$ corrections to the heavy quark limit are given by symmetry breaking terms. The corrections are form of: 
           \begin{multline}
          \mathcal{L}_{1/m_{Q}} = \frac{1}{2m_{Q}}[\lambda_{H} Tr(\overline H_{a}\sigma^{\mu\nu}{H_{a}}\sigma_{\mu\nu}) + \lambda_{S}Tr(\overline S_{a}\sigma^{\mu\nu} S_{a}\sigma_{\mu\nu})\\+\lambda_{T}Tr(\overline T_{a}^{\alpha}\sigma^{\mu\nu}{T_{a}^{\alpha}}\sigma_{\mu\nu})]+\lambda_{X}Tr(\overline X_{a}^{\alpha}\sigma^{\mu\nu}{X_{a}^{\alpha}}\sigma_{\mu\nu}) + \lambda_{Y}Tr(\overline Y_{a}^{\alpha\beta}\sigma^{\mu\nu}{Y_{a}^{\alpha\beta}}\sigma_{\mu\nu}) +\lambda_{Z}Tr(\overline Z_{a}^{\alpha\beta}\sigma^{\mu\nu}{Z_{a}^{\alpha\beta}}\sigma_{\mu\nu})+ \\\lambda_{R}Tr(\overline R_{a}^{\alpha\beta\rho}\sigma^{\mu\nu}{R_{a}^{\alpha\beta\rho}}\sigma_{\mu\nu})]
          \end{multline} 
          Here parameters $\lambda_{H}$, $\lambda_{S}$, $\lambda_{T}$, $\lambda_{X}$, $\lambda_{Y}$, $\lambda_{Z}$, $\lambda_{R}$ are analogous with hyperfine splittings and defined as in Eq. \eqref{hyperpara}. This mass terms in lagrangian represent only first order in $1/m_{Q}$ terms, but higher order terms may also be present otherwise. We are limiting to the first order corrections in $1/m_{Q}$.
 \begin{gather}
     \lambda_{H} = \frac{1}{8}(M^{2}_{P^{*}} - M^{2}_{P}) \\
     \lambda_{S} = \frac{1}{8}({M^{2}_{P_1^{'}}} - {M^{2}_{P_0^*}})\\
     \lambda_{T}=\frac{3}{16}({M^{2}_{P_2^*}}-{M^{2}_{P_1}})\\
     \lambda_{X}=\frac{3}{16}({M^{2}_{P_2}}-{M^{2}_{P_1^*}})\\
      \lambda_{Y}=\frac{5}{24}({M^{2}_{P_3}}-{M^{2}_{P_2^{'*}}})\\
      \lambda_{Z}=\frac{5}{24}({M^{2}_{P_3^*}}-{M^{2}_{P_2^{'}}})\\
      \lambda_{R}=\frac{7}{32}({M^{2}_{P_4^*}}-{M^{2}_{P_3^{'*}}})
     \label{hyperpara}
 \end{gather}
  Here we are motivated by fact that at scale of 1 GeV, when we study HQET, flavour symmetry spontaneously arises for b (bottom quark) and c (charm quark) and hence elegance of flavor symmetry refers to
        \begin{align}
           \label{1eu_eqn}
           \Delta_{F}^{(c)} =\Delta_{F}^{(b)}\\
   \lambda_{F}^{(c)} = \lambda_{F}^{(b)}
   \label{2eu_eqn}
\end{align}

The decays $F\rightarrow  H + M$ (F = H, S, T, X, Y, Z, R
, M presents a light pseudoscalar meson) can be described by effective Lagrangians explained in terms of the fields introduced in (9-14) that valid at leading order in the heavy quark mass and light meson momentum expansion:
\begin{gather}
\label{eq:lagrangian1}
L_{HH}=g_{HH}Tr\{\overline{H}_{a}
H_{b}\gamma_{\mu}\gamma_{5}A^{\mu}_{ba}\}\\
L_{TH}=\frac{g_{TH}}{\Lambda}Tr\{\overline{H}_{a}T^{\mu}_{b}(iD_{\mu}\slashed
A + i\slashed D A_{\mu})_{ba}\gamma_{5}\}+h.c.\\
L_{XH}=\frac{g_{XH}}{\Lambda}Tr\{\overline{H}_{a}X^{\mu}_{b}(iD_{\mu}\slashed
A + i\slashed D A_{\mu})_{ba}\gamma_{5}\}+h.c.
\end{gather}
\begin{multline}
L_{YH}=\frac{1}{\Lambda^{2}}Tr\{\overline{H}_{a}Y^{\mu\nu}_{b}[k^{Y}_{1}\{D_{\mu}
,D_{\nu}\}A_{\lambda}+k^{Y}_{2}(D_{\mu}D_{\lambda}A_{\nu}
+D_{\nu}D_{\lambda}A_{\mu})]_{ba}\gamma^{\lambda}\gamma_{5}\}+h.c.
\end{multline}
\begin{multline}
L_{ZH}=\frac{1}{\Lambda^{2}}Tr\{\overline{H}_{a}Z^{\mu\nu}_{b}[k^{Z}_{1}\{D_{\mu}
,D_{\nu}\}A_{\lambda}+
k^{Z}_{2}(D_{\mu}D_{\lambda}A_{\nu}+D_{\nu}D_{\lambda}A_{\mu})]_{ba}\gamma^{\lambda}\gamma_{5}\}+h.c.
\end{multline}
\begin{multline}
L_{RH}=\frac{1}{\Lambda^{3}}Tr\{\overline{H}_{a}R^{\mu\nu\rho}_{b}[k^{R}_{1}\{D_{\mu}
,D_{\nu}D_{\rho}\}A_{\lambda}+k^{R}_{2}(\{D_{\mu},D_{\rho}\}D_{\lambda}A_{\nu}+\{D_{\nu},D_{\rho}\}D_{\lambda}A_{\mu}\\
+\{D_{\mu},D_{\nu}\}D_{\lambda}A_{\rho})]_{ba}\gamma^{\lambda}\gamma_{5}\}+h.c.
\end{multline}
In these equations $D_{\mu} =
\partial_{\mu}+V_{\mu}$,  $\{D_{\mu},D_{\nu}\}
= D_{\mu}D_{\nu}+D_{\nu}D_{\mu}$ and $\{D_{\mu} ,D_{\nu}D_{\rho}\} =
D_{\mu}D_{\nu}D_{\rho}+D_{\mu}D_{\rho}D_{\nu}+D_{\nu}D_{\mu}D_{\rho}+D_{\nu}D_{\rho}D_{\mu}+D_{\rho}D_{\mu}
D_{\nu}+D_{\rho}D_{\nu}D_{\mu}$. $\Lambda$ is the chiral symmetry
breaking scale taken as 1 GeV. $g_{HH}$, $g_{SH}$, $g_{TH}$, $g_{YH}
= k^{Y}_{1}+k^{Y}_{2}$ and $g_{ZH} = k^{Z}_{1}+k^{Z}_{2}$ are the
strong coupling constants involved. The above equations describe the
interactions of higher excited charm states to the ground state
positive and negative parity charm mesons along with the emission of
light pseudo-scalar mesons $(\pi,\eta,K)$. Using the lagrangians
$L_{HH}, L_{SH}, L_{TH}, L_{YH}, L_{ZH}, L_{RH}$, the two body strong decays
of $Q\overline{q}$ heavy-light charm mesons are given as
$(2^{+},3^{+}) \rightarrow (0^{-},1^{-}) + M$
\begin{gather}
\label{eq:lagrangian} \Gamma(2^{+} \rightarrow 1^{-})=
C_{M}\frac{8g_{ZH}^{2}}{75\pi f_{\pi}^{2}\Lambda^{4}}
\frac{M_{f}}{M_{i}}[p_{M}^{5}(m_{M}^{2}+p_{M}^{2})]\\
\Gamma(2^{+} \rightarrow 0^{-})= C_{M}\frac{4g_{ZH}^{2}}{25\pi f_{\pi}^{2}\Lambda^{4}}\frac{M_{f}}{M_{i}}[p_{M}^{5}(m_{M}^{2}+p_{M}^{2})]\\
\Gamma(3^{+} \rightarrow 1^{-})= C_{M}\frac{4g_{ZH}^{2}}{15\pi
f_{\pi}^{2}\Lambda^{4}}
\frac{M_{f}}{M_{i}}[p_{M}^{5}(m_{M}^{2}+p_{M}^{2})]
\end{gather}
$(3^{+},4^{+}) \rightarrow (0^{-},1^{-}) + M$
\begin{gather}
\label{eq:lagrangian} \Gamma(3^{+} \rightarrow 1^{-})=
C_{M}\frac{36g_{RH}^{2}}{35\pi f_{\pi}^{2}\Lambda^{6}}
\frac{M_{f}}{M_{i}}[p_{M}^{9}]\\
\Gamma(4^{+} \rightarrow 1^{-})= C_{M}\frac{4g_{RH}^{2}}{7\pi f_{\pi}^{2}\Lambda^{6}}\frac{M_{f}}{M_{i}}[p_{M}^{9}]\\
\Gamma(4^{+} \rightarrow 0^{-})= C_{M}\frac{16g_{RH}^{2}}{35\pi
f_{\pi}^{2}\Lambda^{6}}
\frac{M_{f}}{M_{i}}[p_{M}^{9}]
\end{gather}
where $M_{i}$, $M_{f}$ represents initial and final momentum, $\Lambda$ is chiral symmetry breaking scale of 1 GeV. $p_{M}$, $m_{M}$ denotes to final momentum and mass of light pseudoscalar meson. Coupling constant plays key role in phenomenology study of heavy light mesons. These dimensionless coupling constants describes strength of transition between H-H field (negative-negative parity), S-H field (positive-negative parity), T-H field (positive-negative parity).
The coefficient $C_{M}$ for different pseudoscalar particles are:
$C_{\pi^{\pm}}$, $C_{K^{\pm}}$, $C_{K^{0}}$, $C_{\overline{K}^{0}}=1$, $C_{\pi^{0}}=\frac{1}{2}$ and $C_{\eta}=\frac{2}{3}(c\bar{u}, c\bar{d})$ or $\frac{1}{6}(c\bar{s})$. In our paper, we are not including higher order corrections of $\frac{1}{m_{Q}}$ to bring new couplings. We also expect that higher corrections give negligible contribution in comparison of leading order contributions.
\section{Results and Discussions}
Recent discoveries, such as the states $B_{J}(5840)^{0,\pm}$, $B_{J}(5960)$, $B_{J}(5970)$ along with the strange partners $B_{s2}^{*}(5840)$, $B_{sJ}(6064)$, and $B_{sJ}(6114)$, have significantly broadened the established bottom meson spectrum \cite{PDG2024}. Nevertheless, the current experimental exploration on higher excitations in both the bottom and bottom-strange families remains rather scarce. This limitation highlights the need for more comprehensive theoretical investigations to clarify their structure, quantum assignments, and underlying dynamics.

In this work, we conduct a detailed investigation of the higher bottom and bottom-strange mesons using the Heavy Quark Effective Theory (HQET) framework. Specifically, we analyze the $n=2$ radial excited $F$-wave bottom meson states. Our study is organized into two steps: first, we compute the mass spectrum of these mesons, including their strange partners; second, we use the obtained masses as input to predict their strong decay widths and extract the corresponding coupling constants. This strategy provides a means to test the consistency and predictive power of HQET in describing the properties of higher excited bottom and bottom-strange meson states.

\subsection{Masses}
Mass is a key intrinsic property in the spectroscopy of heavy–light mesons, providing information about their internal quark dynamics and interactions. The input values employed in the calculation of radially excited F-wave bottom and bottom-strange meson masses are listed in Table \ref{tab:inputforbottom}

\begin{table}[ht]
\centering
\caption{Input values are taken from Ref. \cite{godfrey2016}. All values are in units of $MeV$.}
\begin{tabular}{|c|c|c|c|c|c|}
\hline
State & $J^{P}$ & $ c\overline{q}$ & $ c\overline{s}$ & $ b\overline{q}$ & $ b\overline{s}$ \\
\hline
 $2^{3}F_{2}$ & $2^{+}$ & 3490 & 3562 & - & -  \\
\hline
$2F_{3}$ & $3^{+}$ & 3498 & 3569 & - & - \\
\hline
$2F^{'}_{3}$ & $3^{+}$ & 3461 & 3540 & - & -  \\
\hline
$2^{3}F_{4}$ & $4^{+}$ & 3466 & 3544 & - & -  \\
\hline
$2^{1}S_{0}$ & $0^{-}$ & 2581 &2673& 5904 &  5984   \\
\hline
$2^{3}S_{1}$ & $1^{-}$ & 2642 &  2732& 5933 & 6012  \\
\hline
\end{tabular}
\label{tab:inputforbottom}
\end{table}

Before evaluating the masses of the $n = 2$, F-wave bottom mesons, 
We first determine the corresponding spin-averaged mass values 
\(\overline{M_{\Tilde{H}}}\), \(\overline{M_{\Tilde{Z}}}\), and \(\overline{M_{\Tilde{R}}}\), 
where the tilde (\(\Tilde{}\)) denotes the radially excited (\( n = 2 \)) states. 
These averages are obtained from the charm meson data listed in Table~\ref{tab:inputforbottom}, 
as defined in Eqs.~\ref{averagemassequation}. 
Subsequently, the Heavy Quark Symmetry (HQS) parameters 
\(\Delta_{\Tilde{H}}\), \(\Delta_{\Tilde{Z}}\), \(\Delta_{\Tilde{R}}\), 
\(\lambda_{\Tilde{H}}\), \(\lambda_{\Tilde{Z}}\), and \(\lambda_{\Tilde{R}}\) 
are evaluated for the same charm meson states according to the definitions given in Eqs.~\ref{lambdaparameters}.

The calculation of 
$n = 2$, $F$ -wave bottom meson masses is preceded by finding of their average mass values $ \overline{M_{\Tilde{H}}}$, $ \overline{M_{\Tilde{Z}}}$, $ \overline{M_{\Tilde{R}}}$  ( $\sim$ labels for $n = 2$) defined in Eqs.\ref{averagemassequation} for charm meson states from Table \ref{tab:inputforbottom}, then HQS (heavy symmetry parameters) $\Delta_{\Tilde{H}}$, $\Delta_{\Tilde{Z}}$, $\Delta_{\Tilde{R}}$, $\lambda_{\Tilde{H}}$, $\lambda_{\Tilde{Z}}$, and  $\lambda_{\Tilde{R}}$ described in Eqs.\ref{lambdaparameters} are computed for same charm meson states. In HQET, $\Delta_{F}$, $\lambda_{F}$ are  assumed to be flavor independent, leading to relations $\Delta_{F}^{(b)} =\Delta_{F}^{(c)}$, $\lambda_{F}^{(b)} = \lambda_{F}^{(c)}$. Using the calculated symmetry parameters  $\Delta_{F}$, $\lambda_{F}$ for charm meson states and then applying heavy quark symmetry, we estimated the masses for $ n = 2$, $F$-wave bottom mesons presented in Table \ref{massresultsforbottommesons}. To illustrate the methodology, we present the detailed calculation of the mass for $B(2^3 F_{2})$ state. From Table \ref{tab:inputforbottom}, using charm states, we calculated $ \overline{M^{c}_{\Tilde{H}}} = 2627.5$ $MeV$, $ \overline{M^{c}_{\Tilde{Z}}} = 3494.06$ $MeV$. Then, using these two values, we have $\Delta^{c}_{\Tilde{Z}} = \overline{M^{c}_{\Tilde{Z}}} - \overline{M^{c}_{\Tilde{H}}} =  866.56$ $MeV$. Using Eq. \ref{lambdaparameters}, we get $\lambda^{c}_{\Tilde{Z}} = 1146.66 $ $MeV^{2}$. The symmetry of these parameters given by Eq. \ref{parametersymmetry} implies that $\Delta^{b}_{\Tilde{Z}} = 866.56$ $MeV$, $\lambda^{b}_{\Tilde{Z}} = 1146.66 $ $MeV^{2}$. Using the values of $\Delta^{b}_{\Tilde{Z}} = 866.56$ $MeV$ and $\lambda^{b}_{\Tilde{Z}} = 1146.66 $ $MeV^{2}$, we obtained the mass of $B(2^3 F_{2}) = 6779.11$ $MeV$. Similarly, by applying the same procedure, we estimated other masses for $n = 2$, $F$-wave bottom mesons presented in Table \ref{massresultsforbottommesons}. 

On comparison, our calculated masses for $n = 2$ $F$-wave bottom meson states are in good agreement with other theoretical estimates for both strange and non-strange states. Our results nicely agree with masses obtained by the relativistic quark model \cite{godfrey2016b} with a deviation of $\pm 1.5\%$. On comparing with Ref. \cite{Jakhad2025}, our results are deviated by $\pm 2.5\%$.

\begin{table}[ht!]
\centering

\caption{Predicted masses for radially excited $F$-wave bottom mesons.}
 \begin{tabular}{| c | c | c | c | c | c | c |}
      \hline
      \multicolumn{1}{|c|}{} & \multicolumn{6}{c|}{Masses of $ 2F$ bottom Mesons ($MeV$)}\\
       \cline{2-7}
     
       \multicolumn{1}{|c|}{$J^{P}(n^{2S+1}L_{J})$}&\multicolumn{3}{c|}{Non-Strange}&\multicolumn{3}{c|}{Strange}\\
       \cline{2-7}
       & \multicolumn{1}{c|}{Present Results}&\multicolumn{1}{c|}{\cite{godfrey2016b}}&\multicolumn{1}{c|}{\cite{Jakhad2025}}&\multicolumn{1}{c|}{Present Results}&\multicolumn{1}{c|}{\cite{godfrey2016b}}&\multicolumn{1}{c|}{\cite{Jakhad2025}}\\
       \hline
        $2^{+}(2^3F_2)$ &  6779.11 &  6704 & 6621& 6851.71 & 6768 & 6767\\
        \hline
        $3^{+}(2F_3)$ &  6801.74 &  6711 &6626  & 6885.35 & 6775 & 6771 \\
        \hline
        $3^{+}(2F^{'}_3)$ &  6760.62& 6673  & 6581& 6795.97 & 6775 & 6741\\
        \hline
        $4^{+}(2^3F_4)$ &  6763.18 &  6679 &  6584& 6856.47 & 6748 & 6744\\
        \hline
   \end{tabular}
   \label{massresultsforbottommesons}
   \end{table}
\subsection{ Upper Bounds of decay widths}
Using obtained masses, we investigated strong decay modes of radially excited F-wave bottom meson states and estimated the upper bounds of the associated decay widths. The strong decays of excited mesons result in the emission of $\pi, K,\eta$, which are regarded as Nambu-Goldstone bosons. Consequently, it is more convenient to examine these interactions. In this work, we restrict our analysis to strong decays of excited heavy-light mesons via emission of pseudoscalar mesons. The decays with emissions of vector mesons ($\omega$, $\rho$, $K^{*}$, $\phi$) are also possible and studied in Ref.\cite{casalbuoni1992, schechter1993, campanella2018}. The contribution of decays with the emission of vector mesons to total decay widths is substantial for these mesonic states. The formulation for decay widths is outlined in Section 2. We apply the effective Lagrangian approach discussed in Section 2 to calculate OZI-allowed two body strong decay widths in terms of associated couplings.

Input values used for computing decay width are $M_{\pi^{0}}$ = 134.97 $MeV$, $M_{\pi^{+}}$ = 139.57 $MeV$, $M_{K^{+}}$ = 493.67 $MeV$, $M_{\eta^{0}}$ = 547.85 $MeV$, $M_{K^{0}}$ = 497.61 $MeV$, $M_{B^{0}}$ = 5279.63 $MeV$, $M_{B^{*}}$ = 5324.65 $MeV$, $M_{B^{0}_{s}}$ = 5366.89 $MeV$, $M_{B^{*}_{s}}$ = 5415.40 $MeV$, and calculated masses for $n = 2$ $F$- wave bottom mesons states.

The computed strong decay widths in terms of $\tilde{g}_{ZH}$, $\tilde{g}_{RH}$ for radially excited $n = 2$ F-wave bottom mesons are presented in Table \ref{table1} and Table \ref{table2} respectively. We also include suppression factors in decays, which arise due to the violation of isospin symmetry in decays. When the mass difference between the parent heavy-light meson with strangeness and the daughter non-strange heavy-light meson is less than the mass of kaons, then the breaking of isospin symmetry takes place, and the suppression factor $\epsilon$ is incorporated in the associated decay mode \cite{gross1979}. To account for isospin violation, the suppression factor is given by :
\begin{align}
    \epsilon^{2} = \frac{3}{16}\left(\frac{m_{d} - m_{u}}{m_{s}-(\frac{m_{u}+m_{d}}{2})}\right) \approx 10^{-4}
\end{align}
Here $m_{u}, m_{d}$, and $m_{s}$ are current quark masses. This suppression factor $\epsilon$ is multiplied with decay modes, which occur with isospin violation.

 \begin{table*}{\normalsize
\renewcommand{\arraystretch}{1.0}
\tabcolsep 0.2cm \caption{\label{table1} Decay width of obtained masses for $2F$ non strange bottom mesons.}
 \centering
   \begin{tabular}{|c|c|c|c|c|c|}
    \hline
    States & $J^{P}$ & Decay Modes& Decay widths& Upper bound of decay widths \\
    \hline
    $B(6779.11)$ & $2^{+}$ & $B^{*+}\pi^-$ & 9319.45$\tilde{g}^{2}_{ZH}$&\\
    & & $B^{*0}\pi^{0}$ & 4666.06$\tilde{g}^{2}_{ZH}$ &\\
    & & $B^{*0}\eta^{0}$&4454.70$\tilde{g}^{2}_{ZH}$&\\
    & & $B^{*}_{s}K^{0}$ & 4665.07$\tilde{g}^{2}_{ZH}$&\\
     & & $B^{+}\pi^{-}$ & 16763.50$\tilde{g}^{2}_{ZH}$&\\
      & & $B^{0}\eta^{0}$ & 8183.26$\tilde{g}^{2}_{ZH}$&\\
       & & $B^{0}_{s}K^{0}$ & 8831.13$\tilde{g}^{2}_{ZH}$&\\
        & & $B^{0}\pi^{0}$ & 8383.30$\tilde{g}^{2}_{ZH}$&\\
    & &Total & 65266.47$\tilde{g}^{2}_{ZH}$ & 319.80\\
    
     \hline
     $B(6478.93)$ & $3^{+}$ & $B^{*0}\pi^0$ &  12829.5$\tilde{g}^{2}_{ZH}$& \\
      & &  $B^{*+}\pi^-$ & 25629.7$\tilde{g}^{2}_{ZH}$& \\
      & & $B^{*0}\eta^0$ & 12387.8$\tilde{g}^{2}_{ZH}$& \\
       & & $B_{s}^{*0}K^0$ & 13063.2$\tilde{g}^{2}_{ZH}$& \\
          & &Total & 59606.73$\tilde{g}^{2}_{ZH}$& 292.07  \\
         
    \hline
   $B(6447.76)$ & $3^{+}$ & $B^{*0}\pi^0$ &  67122$\tilde{g}^{2}_{RH}$& \\
      & &  $B^{*+}\pi^-$ & 133857$\tilde{g}^{2}_{RH}$& \\
      & & $B^{*0}\eta^0$ & 45487$\tilde{g}^{2}_{RH}$& \\
       & & $B_{s}^{*0}K^0$ & 43279.8$\tilde{g}^{2}_{RH}$& \\
          & &Total &289745.8$\tilde{g}^{2}_{RH}$&  1043.80\\
        
     \hline
      $B(6450.14)$ & $4^{+}$ & $B^{*+}\pi^-$ & 75438.8$\tilde{g}^{2}_{RH}$&\\
    & & $B^{*0}\pi^{0}$ & 37825.65$\tilde{g}^{2}_{RH}$ &\\
    & & $B^{*0}\eta^{0}$&25701.73$\tilde{g}^{2}_{RH}$&\\
    & & $B^{*}_{s}K^{0}$ & 24479.4$\tilde{g}^{2}_{RH}$&\\
     & & $B^{+}\pi^{-}$ & 76713.4$\tilde{g}^{2}_{RH}$&\\
     & & $B^{0}\eta^{0}$ & 27237.53$\tilde{g}^{2}_{RH}$&\\
       & & $B^{0}_{s}K^{0}$ & 27009.8$\tilde{g}^{2}_{RH}$&\\
        & & $B^{0}\pi^{0}$ & 38400.85$\tilde{g}^{2}_{RH}$&\\
    & & Total & 332807.16$\tilde{g}^{2}_{RH}$ &1198.10\\
    
          \hline
    \end{tabular}
    }
    \end{table*}
   \begin{table*}{\normalsize
\renewcommand{\arraystretch}{1.0}
\tabcolsep 0.2cm \caption{\label{table2} Decay width of obtained masses for $2F$ strange bottom mesons.}
 \centering
   \begin{tabular}{|c|c|c|c|c|c|}
    \hline
    States & $J^{P}$ & Decay Modes& Decay Widths& Upper bound of decay widths \\
    \hline
    $B_{s}(6518.28)$ & $2^{+}$ & $B^{*0}K^0$ & 9900.15$\tilde{g}^{2}_{ZH}$&\\
    & & $B^{*-}K^{+}$ & 9945.26$\tilde{g}^{2}_{ZH}$ &\\
    & & $B_{s}^{*0}\eta^{0}$&1036.02$\tilde{g}^{2}_{ZH}$&\\
    & & $B^{*}_{s}\pi^{0}$ & 4383.33 $\tilde{g}^{2}_{ZH} \times 10^{-4}$&\\
     & & $B^{0}K^{0}$ & 17875.8$\tilde{g}^{2}_{ZH}$&\\
      & & $B^{-}K^{+}$ & 17968.8$\tilde{g}^{2}_{ZH}$&\\
       & & $B^{0}_{s}\eta^{0}$ & 1940.23$\tilde{g}^{2}_{ZH}$&\\
        & & $B_{s}^{0}\pi^{0}$ & 8008.05$\tilde{g}^{2}_{ZH}\times 10^{-4}$&\\
    & &Total & 58667.49$\tilde{g}^{2}_{ZH} $ & 586.67 \\                    
     \hline
     $B_{s}(6523.21)$ & $3^{+}$ & $B^{*-}K^+$ & 28727.7$\tilde{g}^{2}_{ZH}$& \\
      & &  $B^{*0}K^0$ &28603.5$\tilde{g}^{2}_{ZH}$& \\
      & & $B_{s}^{*0}\eta^0$ &3040.48$\tilde{g}^{2}_{ZH}$& \\
       & & $B_{s}^{*0}\pi^0$ & 12643.15$\tilde{g}^{2}_{ZH}\times 10^{-4}$& \\
          & &Total & 60372.94$\tilde{g}^{2}_{ZH}$ & 603.73 \\
          
    \hline
   $B_{s}(6506.05)$ & $3^{+}$ & $B^{*-}K^+$ &  98339.9$\tilde{g}^{2}_{ZH}$& \\
      & &  $B^{*0}K^0$ & 97405.9$\tilde{g}^{2}_{RH}$& \\
      & & $B_{s}^{*0}\eta^0$ & 9747.15$\tilde{g}^{2}_{RH}$&\\
       & & $B_{s}^{*0}\pi^0$ & 49789.25$\tilde{g}^{2}_{RH}\times 10^{-4}$& \\
          & &Total & 205497.92$\tilde{g}^{2}_{RH}$ & 513.74 \\
        
     \hline
       $B_{s}(6508.01)$ & $4^{+}$ & $B^{*0}K^0$&  77852.3$\tilde{g}^{2}_{RH}$&\\
    & & $B^{*-}K^{+}$ & 78540.9$\tilde{g}^{2}_{RH}$&\\
    & & $B_{s}^{*0}\eta^{0}$& 6656.56$\tilde{g}^{2}_{RH}$&\\
    & & $B^{*}_{s}\pi^{0}$ & 39073.65$\tilde{g}^{2}_{RH}\times 10^{-4}$&\\
     & & $B^{0}K^{0}$ & 79908.2$\tilde{g}^{2}_{RH}$&\\
      & & $B^{-}K^{+}$ & 80673.4$\tilde{g}^{2}_{RH}$&\\
       & & $B^{0}_{s}\eta^{0}$ & 7201.33 $\tilde{g}^{2}_{RH}$&\\
        & & $B_{s}^{0}\pi^{0}$ & 40377.4$\tilde{g}^{2}_{RH}\times 10^{-4}$&\\
    & &Total & 330840.63$\tilde{g}^{2}_{RH}$& 827.10 \\
    
          \hline
    \end{tabular}
    }
    \end{table*} 

The computed strong decay widths in terms of coupling constants $\tilde{g}_{ZH}$, $\tilde{g}_{RH}$ for $n = 2$ F-wave bottom meson states are presented in Table \ref{table1} and Table\ref{table2}, respectively. Without enough experimental data, it is not possible to determine values of decay widths of excited bottom meson states from heavy quark symmetry solely, but the upper bounds to their decay widths are mentioned in Table \ref{table1} and Table \ref{table2}. In our study, we are taking limited modes of decay, and that also only to ground state. We believe that a particular state like $B(65266)$ gives 65266.47 $\tilde{g}_{HH}^2$ total decay width; when computed by adding values of coupling constants in its expression, we provided an upper bound on decay widths of the associated state. Therefore, these upper bounds may provide important information about other associated bottom states. Large fractions of the decay width of any excited state are dominated by modes that include the ground state. This work also provides a lower limit to the total decay width, offering important clues for forthcoming experimental studies. The weak and radiative decays are not included in the computed decay widths of charm and bottom mesons. We also exclude decays via emissions of vector mesons ($\omega,\rho, K^*,\phi$). They give the contribution of $\pm{50}$ $MeV$ \cite{godfrey2016,godfrey2016b} to total decay widths for states analyzed above.

The coupling constant plays an important role in hadron spectroscopy. They are coupled with decays and give information about the strength of strong transitions of excited heavy meson doublets into the highest heavy meson doublets. Here, dimensionless coupling constants $g_{HH}$, $g_{SH}$, $g_{TH}$, $g_{ZH}$, $g_{RH}$ give the strength of transitions between $H-H, S-H, T-H, Z-H$, and $R-H$ fields, respectively. Values of coupling constants are more for ground state transitions ($H-H$ fields) than excited states ($S-H, T-H, X-H, Z-H, R-H$ fields) transition shown by value of $g_{HH} =0.64\pm0.075$ \cite{colangelo2012} while $g_{SH} =0.56\pm 0.04$, $g_{TH} =0.43\pm 0.01$ \cite{colangelo2012}, $g_{XH} = 0.24$ \cite{wang2014}. Also, values of coupling constants are low at higher orders ($n = 2$, $n = 3$) in comparison to lower order ($n = 1$) interactions \cite{colangelo2012,casalbuoni1997} like $\tilde{g}_{HH} = 0.28\pm 0.015$, $\tilde{g}_{SH} = 0.18\pm 0.01$ so on. In the same manner, we assumed that the value of the coupling constants $\tilde{g}_{ZH}$, $\tilde{g}_{RH}$ is less than $g_{ZH}$ and $g_{RH}$. In our previous work, we have calculated the value of coupling constants $g_{ZH}$ and $g_{RH}$. For non-strange bottom mesons, 
$g_{ZH} = 0.07$ and $g_{RH} = 0.06$, while for strange bottom mesons, $g_{ZH} = 0.10$ and $g_{RH} = 0.05$. By inserting these values, we calculated upper bounds of decay widths for associated bottom meson states. These findings may be helpful for future experimental outcomes.

\section{Conculsion}
 Heavy quark symmetry serves as a fundamental framework for describing the spectroscopy of hadrons containing a single heavy quark. Utilizing available experimental and theoretical data on charm mesons, we employ heavy quark symmetry to predict the masses of $n = 2$ F-wave bottom meson states. Based on these computed masses, we investigate the decay widths of excited states transitioning to the ground state via the emission of pseudoscalar mesons. The decay widths are expressed in terms of coupling constants, which are determined by inserting values of coupling constants taken from our previous work. These values provide an upper bound on decay widths, giving valuable insights into the properties of other associated bottom meson states. These findings may assist in the analysis of upcoming experimental results.

\bibliography{ref}

@article{aaij2021,
    author = "Aaij, Roel and others",
    collaboration = "LHCb",
    title = "{Observation of a New Excited $D^+_s$ Meson in $B^0 \rightarrow D^- D^+ K^+ \pi^-$ Decays}",
    eprint = "2011.09112",
    archivePrefix = "arXiv",
    primaryClass = "hep-ex",
    reportNumber = "CERN-EP-2020-203, LHCb-PAPER-2020-034",
    doi = "10.1103/PhysRevLett.126.122002",
    journal = "Phys. Rev. Lett.",
    volume = "126",
    number = "12",
    pages = "122002",
    year = "2021"
}

@article{1,
doi = {10.1088/1361-6633/aca3b6},
url = {https://dx.doi.org/10.1088/1361-6633/aca3b6},
year = {2022},
month = {dec},
publisher = {IOP Publishing},
volume = {86},
number = {2},
pages = {026201},
author = {Hua-Xing Chen and Wei Chen and Xiang Liu and Yan-Rui Liu and Shi-Lin Zhu},
title = {An updated review of the new hadron states},
journal = {Reports on Progress in Physics},
}

@article{2,
  title = {Mass spectra and strong decays of charmed and charmed-strange mesons},
  author = {Ni, Ru-Hui and Li, Qi and Zhong, Xian-Hui},
  journal = {Phys. Rev. D},
  volume = {105},
  issue = {5},
  pages = {056006},
  numpages = {39},
  year = {2022},
  month = {Mar},
  publisher = {American Physical Society},
  doi = {10.1103/PhysRevD.105.056006},
  url = {https://link.aps.org/doi/10.1103/PhysRevD.105.056006}
}

@article{4,
title = {The Ds0(2590)+ as the dressed cs¯(21S0) meson in a coupled-channels calculation},
journal = {Physics Letters B},
volume = {827},
pages = {136998},
year = {2022},
issn = {0370-2693},
doi = {https://doi.org/10.1016/j.physletb.2022.136998},
url = {https://www.sciencedirect.com/science/article/pii/S0370269322001320},
author = {Pablo G. Ortega and Jorge Segovia and David R. Entem and Francisco Fernández},
}

@article{5,
  title = {Observation of new resonances decaying to $D\ensuremath{\pi}$ and ${D}^{*}\ensuremath{\pi}$ in inclusive ${e}^{+}{e}^{\ensuremath{-}}$ collisions near $\sqrt{s}=10.58\text{ }\text{ }\mathrm{GeV}$},
  author = {del Amo Sanchez, P. and Lees, J. P. and Poireau, V. and Prencipe, E. and Tisserand, V. and Garra Tico, J. and Grauges, E. and Martinelli, M. and Palano, A. and Pappagallo, M. and Eigen, G. and Stugu, B. and Sun, L. and Battaglia, M. and Brown, D. N. and Hooberman, B. and Kerth, L. T. and Kolomensky, Yu. G. and Lynch, G. and Osipenkov, I. L. and Tanabe, T. and Hawkes, C. M. and Watson, A. T. and Koch, H. and Schroeder, T. and Asgeirsson, D. J. and Hearty, C. and Mattison, T. S. and McKenna, J. A. and Khan, A. and Randle-Conde, A. and Blinov, V. E. and Buzykaev, A. R. and Druzhinin, V. P. and Golubev, V. B. and Onuchin, A. P. and Serednyakov, S. I. and Skovpen, Yu. I. and Solodov, E. P. and Todyshev, K. Yu. and Yushkov, A. N. and Bondioli, M. and Curry, S. and Kirkby, D. and Lankford, A. J. and Mandelkern, M. and Martin, E. C. and Stoker, D. P. and Atmacan, H. and Gary, J. W. and Liu, F. and Long, O. and Vitug, G. M. and Campagnari, C. and Hong, T. M. and Kovalskyi, D. and Richman, J. D. and West, C. and Eisner, A. M. and Heusch, C. A. and Kroseberg, J. and Lockman, W. S. and Martinez, A. J. and Schalk, T. and Schumm, B. A. and Seiden, A. and Winstrom, L. O. and Cheng, C. H. and Doll, D. A. and Echenard, B. and Hitlin, D. G. and Ongmongkolkul, P. and Porter, F. C. and Rakitin, A. Y. and Andreassen, R. and Dubrovin, M. S. and Mancinelli, G. and Meadows, B. T. and Sokoloff, M. D. and Bloom, P. C. and Ford, W. T. and Gaz, A. and Nagel, M. and Nauenberg, U. and Smith, J. G. and Wagner, S. R. and Ayad, R. and Toki, W. H. and Jasper, H. and Karbach, T. M. and Merkel, J. and Petzold, A. and Spaan, B. and Wacker, K. and Kobel, M. J. and Schubert, K. R. and Schwierz, R. and Bernard, D. and Verderi, M. and Clark, P. J. and Playfer, S. and Watson, J. E. and Andreotti, M. and Bettoni, D. and Bozzi, C. and Calabrese, R. and Cecchi, A. and Cibinetto, G. and Fioravanti, E. and Franchini, P. and Luppi, E. and Munerato, M. and Negrini, M. and Petrella, A. and Piemontese, L. and Baldini-Ferroli, R. and Calcaterra, A. and de Sangro, R. and Finocchiaro, G. and Nicolaci, M. and Pacetti, S. and Patteri, P. and Peruzzi, I. M. and Piccolo, M. and Rama, M. and Zallo, A. and Contri, R. and Guido, E. and Lo Vetere, M. and Monge, M. R. and Passaggio, S. and Patrignani, C. and Robutti, E. and Tosi, S. and Bhuyan, B. and Prasad, V. and Lee, C. L. and Morii, M. and Adametz, A. and Marks, J. and Uwer, U. and Bernlochner, F. U. and Ebert, M. and Lacker, H. M. and Lueck, T. and Volk, A. and Dauncey, P. D. and Tibbetts, M. and Behera, P. K. and Mallik, U. and Chen, C. and Cochran, J. and Crawley, H. B. and Dong, L. and Meyer, W. T. and Prell, S. and Rosenberg, E. I. and Rubin, A. E. and Gritsan, A. V. and Guo, Z. J. and Arnaud, N. and Davier, M. and Derkach, D. and Firmino da Costa, J. and Grosdidier, G. and Le Diberder, F. and Lutz, A. M. and Malaescu, B. and Perez, A. and Roudeau, P. and Schune, M. H. and Serrano, J. and Sordini, V. and Stocchi, A. and Wang, L. and Wormser, G. and Lange, D. J. and Wright, D. M. and Bingham, I. and Chavez, C. A. and Coleman, J. P. and Fry, J. R. and Gabathuler, E. and Gamet, R. and Hutchcroft, D. E. and Payne, D. J. and Touramanis, C. and Bevan, A. J. and Di Lodovico, F. and Sacco, R. and Sigamani, M. and Cowan, G. and Paramesvaran, S. and Wren, A. C. and Brown, D. N. and Davis, C. L. and Denig, A. G. and Fritsch, M. and Gradl, W. and Hafner, A. and Alwyn, K. E. and Bailey, D. and Barlow, R. J. and Jackson, G. and Lafferty, G. D. and Anderson, J. and Cenci, R. and Jawahery, A. and Roberts, D. A. and Simi, G. and Tuggle, J. M. and Dallapiccola, C. and Salvati, E. and Cowan, R. and Dujmic, D. and Sciolla, G. and Zhao, M. and Lindemann, D. and Patel, P. M. and Robertson, S. H. and Schram, M. and Biassoni, P. and Lazzaro, A. and Lombardo, V. and Palombo, F. and Stracka, S. and Cremaldi, L. and Godang, R. and Kroeger, R. and Sonnek, P. and Summers, D. J. and Nguyen, X. and Simard, M. and Taras, P. and De Nardo, G. and Monorchio, D. and Onorato, G. and Sciacca, C. and Raven, G. and Snoek, H. L. and Jessop, C. P. and Knoepfel, K. J. and LoSecco, J. M. and Wang, W. F. and Corwin, L. A. and Honscheid, K. and Kass, R. and Morris, J. P. and Blount, N. L. and Brau, J. and Frey, R. and Igonkina, O. and Kolb, J. A. and Rahmat, R. and Sinev, N. B. and Strom, D. and Strube, J. and Torrence, E. and Castelli, G. and Feltresi, E. and Gagliardi, N. and Margoni, M. and Morandin, M. and Posocco, M. and Rotondo, M. and Simonetto, F. and Stroili, R. and Ben-Haim, E. and Bonneaud, G. R. and Briand, H. and Calderini, G. and Chauveau, J. and Hamon, O. and Leruste, Ph. and Marchiori, G. and Ocariz, J. and Prendki, J. and Sitt, S. and Biasini, M. and Manoni, E. and Rossi, A. and Angelini, C. and Batignani, G. and Bettarini, S. and Carpinelli, M. and Casarosa, G. and Cervelli, A. and Forti, F. and Giorgi, M. A. and Lusiani, A. and Neri, N. and Paoloni, E. and Rizzo, G. and Walsh, J. J. and Lopes Pegna, D. and Lu, C. and Olsen, J. and Smith, A. J. S. and Telnov, A. V. and Anulli, F. and Baracchini, E. and Cavoto, G. and Faccini, R. and Ferrarotto, F. and Ferroni, F. and Gaspero, M. and Li Gioi, L. and Mazzoni, M. A. and Piredda, G. and Renga, F. and Hartmann, T. and Leddig, T. and Schr\"oder, H. and Waldi, R. and Adye, T. and Franek, B. and Olaiya, E. O. and Wilson, F. F. and Emery, S. and Hamel de Monchenault, G. and Vasseur, G. and Y\`eche, Ch. and Zito, M. and Allen, M. T. and Aston, D. and Bard, D. J. and Bartoldus, R. and Benitez, J. F. and Cartaro, C. and Convery, M. R. and Dorfan, J. and Dubois-Felsmann, G. P. and Dunwoodie, W. and Field, R. C. and Franco Sevilla, M. and Fulsom, B. G. and Gabareen, A. M. and Graham, M. T. and Grenier, P. and Hast, C. and Innes, W. R. and Kelsey, M. H. and Kim, H. and Kim, P. and Kocian, M. L. and Leith, D. W. G. S. and Li, S. and Lindquist, B. and Luitz, S. and Luth, V. and Lynch, H. L. and MacFarlane, D. B. and Marsiske, H. and Muller, D. R. and Neal, H. and Nelson, S. and O'Grady, C. P. and Ofte, I. and Perl, M. and Pulliam, T. and Ratcliff, B. N. and Roodman, A. and Salnikov, A. A. and Santoro, V. and Schindler, R. H. and Schwiening, J. and Snyder, A. and Su, D. and Sullivan, M. K. and Sun, S. and Suzuki, K. and Thompson, J. M. and Va'vra, J. and Wagner, A. P. and Weaver, M. and West, C. A. and Wisniewski, W. J. and Wittgen, M. and Wright, D. H. and Wulsin, H. W. and Yarritu, A. K. and Young, C. C. and Ziegler, V. and Chen, X. R. and Park, W. and Purohit, M. V. and White, R. M. and Wilson, J. R. and Sekula, S. J. and Bellis, M. and Burchat, P. R. and Edwards, A. J. and Miyashita, T. S. and Ahmed, S. and Alam, M. S. and Ernst, J. A. and Pan, B. and Saeed, M. A. and Zain, S. B. and Guttman, N. and Soffer, A. and Lund, P. and Spanier, S. M. and Eckmann, R. and Ritchie, J. L. and Ruland, A. M. and Schilling, C. J. and Schwitters, R. F. and Wray, B. C. and Izen, J. M. and Lou, X. C. and Bianchi, F. and Gamba, D. and Pelliccioni, M. and Bomben, M. and Lanceri, L. and Vitale, L. and Lopez-March, N. and Martinez-Vidal, F. and Milanes, D. A. and Oyanguren, A. and Albert, J. and Banerjee, Sw. and Choi, H. H. F. and Hamano, K. and King, G. J. and Kowalewski, R. and Lewczuk, M. J. and Nugent, I. M. and Roney, J. M. and Sobie, R. J. and Gershon, T. J. and Harrison, P. F. and Latham, T. E. and Puccio, E. M. T. and Band, H. R. and Dasu, S. and Flood, K. T. and Pan, Y. and Prepost, R. and Vuosalo, C. O. and Wu, S. L.},
  collaboration = {BABAR Collaboration},
  journal = {Phys. Rev. D},
  volume = {82},
  issue = {11},
  pages = {111101},
  numpages = {9},
  year = {2010},
  month = {Dec},
  publisher = {American Physical Society},
  doi = {10.1103/PhysRevD.82.111101},
  url = {https://link.aps.org/doi/10.1103/PhysRevD.82.111101}
}

@article{6,
  title={Study of $DJ$ meson decays to $D^+ \pi^-$, $D^0\pi^+$ and $D^{∗+} \pi^-$ final states in pp collisions},
  author={Aaij, R and Adeva, B and Adinolfi, M and Adrover, C and Affolder, A and Ajaltouni, Ziad and Albrecht, J and Alessio, F and Alexander, M and Ali, S and others},
  journal={Journal of High Energy Physics},
  volume={2013},
  number={9},
  pages={1-26},
  year={2013},
  publisher={Springer}
}

@article{7,
  title={First observation and amplitude analysis of the $B^- \rightarrow D^+ K^- \pi^-$ decay},
  author={Aaij, R and Adeva, B and Adinolfi, M and Affolder, A and Ajaltouni, Z and Akar, S and Albrecht, J and Alessio, F and Alexander, M and Ali, S and others},
  journal={Physical Review D},
  volume={91},
  number={9},
  pages={092002},
  year={2015},
  publisher={APS}
}

@article{8,
  title={Observation of a New $D_s$ Meson Decaying to $D K$ at a Mass of 2.86 $GeV/c^2$},
  author={Aubert, Bernard and Barate, R and Bona, M and Boutigny, D and Couderc, F and Karyotakis, Y and Lees, JP and Poireau, V and Tisserand, V and Zghiche, A and others},
  journal={Physical review letters},
  volume={97},
  number={22},
  pages={222001},
  year={2006},
  publisher={APS}
}

@article{9,
  title={Observation of a New $D_{sJ}$ Meson in $B^+ \rightarrow \bar D^0 D^0 K^+$ Decays},
  author={Brodzicka, J and Palka, H and Adachi, I and Aihara, H and Aulchenko, V and Bakich, AM and Barberio, E and Bay, A and Bedny, I and Bitenc, U and others},
  journal={Physical review letters},
  volume={100},
  number={9},
  pages={092001},
  year={2008},
  publisher={APS}
}

@article{11,
    author = "Workman, R. L. and Others",
    collaboration = "Particle Data Group",
    title = "{Review of Particle Physics}",
    doi = "10.1093/ptep/ptac097",
    journal = "PTEP",
    volume = "2022",
    pages = "083C01",
    year = "2022"
}

@article{16,
  title = {Excited heavy-light systems and hadronic transitions},
  author = {Di Pierro, M. and Eichten, E.},
  journal = {Phys. Rev. D},
  volume = {64},
  issue = {11},
  pages = {114004},
  numpages = {20},
  year = {2001},
  month = {Oct},
  publisher = {American Physical Society},
  doi = {10.1103/PhysRevD.64.114004},
  url = {https://link.aps.org/doi/10.1103/PhysRevD.64.114004}
}

@article{24,
  title={Analysis of strong decays of the charmed mesons $D_J$(2580), $D_{J}^*$(2650), $D_J$(2740), $D_{J}^*$(2760), $D_J$ (3000), $D_{J}^*$(3000)},
  author={Wang, Zhi-Gang},
  journal={Physical Review D},
  volume={88},
  number={11},
  pages={114003},
  year={2013},
  publisher={APS}
}

@article{13a,
  title={Excited heavy-light systems and hadronic transitions},
  author={Di Pierro, Massimo and Eichten, Estia},
  journal={Physical Review D},
  volume={64},
  number={11},
  pages={114004},
  year={2001},
  publisher={APS}
}

@article{14a,
  title={Higher bottom and bottom-strange mesons},
  author={Sun, Yuan and Song, Qin-Tao and Chen, Dian-Yong and Liu, Xiang and Zhu, Shi-Lin},
  journal={Physical Review D},
  volume={89},
  number={5},
  pages={054026},
  year={2014},
  publisher={APS}
}

@article{15a,
  title = {$B$ and ${B}_{s}$ meson spectroscopy},
  author = {Godfrey, Stephen and Moats, Kenneth and Swanson, Eric S.},
  journal = {Phys. Rev. D},
  volume = {94},
  issue = {5},
  pages = {054025},
  numpages = {41},
  year = {2016},
  month = {Sep},
  publisher = {American Physical Society},
  doi = {10.1103/PhysRevD.94.054025},
  url = {https://link.aps.org/doi/10.1103/PhysRevD.94.054025}
}

@article{16a,
  title={Excited bottom and bottom-strange mesons in the quark model},
  author={L{\"u}, Qi-Fang and Pan, Ting-Ting and Wang, Yan-Yan and Wang, En and Li, De-Min},
  journal={Physical Review D},
  volume={94},
  number={7},
  pages={074012},
  year={2016},
  publisher={APS}
}

@article{17a,
  title={Decays and spectrum of bottom and bottom strange mesons},
  author={Asghar, Ishrat and Masud, Bilal and Swanson, ES and Akram, Faisal and Atif Sultan, M},
  journal={The European Physical Journal A},
  volume={54},
  pages={1--22},
  year={2018},
  publisher={Springer}
}

@article{18a,
  title={Spectroscopic assignments of the excited $B$-mesons},
  author={Godfrey, Stephen and Moats, Kenneth},
  journal={The European Physical Journal A},
  volume={55},
  number={5},
  pages={84},
  year={2019},
  publisher={Springer}
}

@article{19a,
  title={Strong decays of-wave mixing heavy-light states},
  author={Wang, Zhi-Hui and Zhang, Yi and Wang, Tian-Hong and Jiang, Yue and Li, Qiang and Wang, Guo-Li},
  journal={Chinese Physics C},
  volume={42},
  number={12},
  pages={123101},
  year={2018},
  publisher={IOP Publishing}
}

@article{20a,
  title={Analysis of the excited bottom and bottom-strange states $B_1$(5721), $B_{2}^*$(5747), $B_{s1}$(5830), $B_{s2}^*$(5840), $B_J$ (5840) and $B_J$ (5970) of the $B$ meson family},
  author={Yu, Guo-Liang and Wang, Zhi-Gang},
  journal={Chinese Physics C},
  volume={44},
  number={3},
  pages={033103},
  year={2020},
  publisher={IOP Publishing}
}

@article{21a,
  title={Strong decay constants of heavy tensor mesons in light cone $QCD$ sum rules},
  author={Alhendi, HA and Aliev, TM and Savc{\i}, M},
  journal={Journal of High Energy Physics},
  volume={2016},
  number={4},
  pages={1--16},
  year={2016},
  publisher={Springer}
}

@article{22a,
  title={Open-flavor strong decays of open-charm and open-bottom mesons in the $3P_0$ model},
  author={Ferretti, J and Santopinto, E},
  journal={Physical Review D},
  volume={97},
  number={11},
  pages={114020},
  year={2018},
  publisher={APS}
}

@article{23a,
  title={Strong decay of the heavy tensor mesons with $QCD$ sum rules},
  author={Wang, Zhi-Gang},
  journal={The European Physical Journal C},
  volume={74},
  pages={1--9},
  year={2014},
  publisher={Springer}
}

@article{24a,
  title={Newly observed $B$ (5970) and the predictions of its spin and strange partners},
  author={Xu, Hao and Liu, Xiang and Matsuki, Takayuki},
  journal={Physical Review D},
  volume={89},
  number={9},
  pages={097502},
  year={2014},
  publisher={APS}
}

@article{25a,
  title={Strong decays of the bottom mesons $B_1$ (5721), $B_2$ (5747), $B_{s1}$ (5830), $B_{s2}$ (5840) and $B$ (5970)},
  author={Wang, Zhi-Gang},
  journal={The European Physical Journal Plus},
  volume={129},
  number={8},
  pages={186},
  year={2014},
  publisher={Springer}
}

@article{26a,
  title={Strong decays of the radial excited states $B(2S)$ and $D(2S)$},
  author={Zhang, Jin-Mei and Wang, Guo-Li},
  journal={Physics Letters B},
  volume={684},
  number={4-5},
  pages={221--223},
  year={2010},
  publisher={Elsevier}
}

@article{27a,
  title={$B_{s1}$(5830) and $B_{s2}^*$(5840)},
  author={Luo, Zhi-Gang and Chen, Xiao-Lin and Liu, Xiang and others},
  journal={Physical Review D},
  volume={79},
  number={7},
  pages={074020},
  year={2009},
  publisher={APS}
}

@article{28a,
  title={Placing the newly observed state $B_J$ (5840) in bottom spectra along with states $B_1$ (5721), $B_{2}^*$(5747), $B_{s1}$(5830), $B_{2s}^*$(5840), and $B_J$(5970)},
  author={Gupta, Pallavi and Upadhyay, A},
  journal={Physical Review D},
  volume={99},
  number={9},
  pages={094043},
  year={2019},
  publisher={APS}
}

@article{29a,
  title={The effect of B$\pi$ continuum in the $QCD$ sum rules for the ($0^+, 1^+$) heavy meson doublet in HQET},
  author={Zhu, Shi-Lin and Dai, Yuan-Ben},
  journal={Modern Physics Letters A},
  volume={14},
  number={34},
  pages={2367--2377},
  year={1999},
  publisher={World Scientific}
}

@article{30a,
  title={Strong and electromagnetic decays of excited heavy mesons},
  author={Hiorth {\"O}rsland, A and H{\"o}gaasen, H},
  journal={The European Physical Journal C-Particles and Fields},
  volume={9},
  pages={503--510},
  year={1999},
  publisher={Springer}
}

@article{31a,
  title={Strong coupling constants and radiative decays of the heavy tensor mesons},
  author={Yu, Guo-Liang and Wang, Zhi-Gang and Li, Zhen-Yu},
  journal={The European Physical Journal C},
  volume={79},
  number={9},
  pages={798},
  year={2019},
  publisher={Springer}
}

@article{32a,
  title={New meson spectroscopy with open charm and beauty},
  author={Colangelo, P and De Fazio, F and Giannuzzi, F and Nicotri, S},
  journal={Physical Review D},
  volume={86},
  number={5},
  pages={054024},
  year={2012},
  publisher={APS}
}

@article{33a,
  title={Strong decays of heavy-light mesons in a chiral quark model},
  author={Zhong, Xian-Hui and Zhao, Qiang},
  journal={Physical Review D},
  volume={78},
  number={1},
  pages={014029},
  year={2008},
 publisher={APS}
}

@article{34a,
  title={Study of $B$, $B_s$ mesons using heavy quark effective theory},
  author={Gandhi, Keval and Rai, Ajay Kumar},
  journal={The European Physical Journal C},
  volume={82},
  number={9},
  pages={777},
  year={2022},
  publisher={Springer}
}

@article{wang2014,
    author = "Wang, Zhi-Gang and Huang, Tao",
    title = "{The $Z_b(10610)$ and $Z_b(10650)$ as axial-vector tetraquark states in the QCD sum rules}",
    doi = "10.1016/j.nuclphysa.2014.08.084",
    journal = "\href{https://doi.org/10.1016/j.nuclphysa.2014.08.084}{Nucl. Phys. A}",
    volume = "{930}",
    pages = "{63}",
    year = "{2014}"
}

@article{aaltonen2008,
  title = {Observation of the Decay ${B}_{c}^{\ifmmode\pm\else\textpm\fi{}}\ensuremath{\rightarrow}J/\ensuremath{\psi}{\ensuremath{\pi}}^{\ifmmode\pm\else\textpm\fi{}}$ and Measurement of the ${B}_{c}^{\ifmmode\pm\else\textpm\fi{}}$ Mass},
  author = {Aaltonen, T. and Adelman, J. and Akimoto, T. and Albrow, M. G. and \'Alvarez Gonz\'alez, B. and Amerio, S. and Amidei, D. and Anastassov, A. and Annovi, A. and Antos, J. and Aoki, M. and Apollinari, G. and Apresyan, A. and Arisawa, T. and Artikov, A. and Ashmanskas, W. and Attal, A. and Aurisano, A. and Azfar, F. and Azzi-Bacchetta, P. and Azzurri, P. and Bacchetta, N. and Badgett, W. and Barbaro-Galtieri, A. and Barnes, V. E. and Barnett, B. A. and Baroiant, S. and Bartsch, V. and Bauer, G. and Beauchemin, P.-H. and Bedeschi, F. and Bednar, P. and Behari, S. and Bellettini, G. and Bellinger, J. and Belloni, A. and Benjamin, D. and Beretvas, A. and Beringer, J. and Berry, T. and Bhatti, A. and Binkley, M. and Bisello, D. and Bizjak, I. and Blair, R. E. and Blocker, C. and Blumenfeld, B. and Bocci, A. and Bodek, A. and Boisvert, V. and Bolla, G. and Bolshov, A. and Bortoletto, D. and Boudreau, J. and Boveia, A. and Brau, B. and Bridgeman, A. and Brigliadori, L. and Bromberg, C. and Brubaker, E. and Budagov, J. and Budd, H. S. and Budd, S. and Burkett, K. and Busetto, G. and Bussey, P. and Buzatu, A. and Byrum, K. L. and Cabrera, S. and Campanelli, M. and Campbell, M. and Canelli, F. and Canepa, A. and Carlsmith, D. and Carosi, R. and Carrillo, S. and Carron, S. and Casal, B. and Casarsa, M. and Castro, A. and Catastini, P. and Cauz, D. and Cavalli-Sforza, M. and Cerri, A. and Cerrito, L. and Chang, S. H. and Chen, Y. C. and Chertok, M. and Chiarelli, G. and Chlachidze, G. and Chlebana, F. and Cho, K. and Chokheli, D. and Chou, J. P. and Choudalakis, G. and Chuang, S. H. and Chung, K. and Chung, W. H. and Chung, Y. S. and Ciobanu, C. I. and Ciocci, M. A. and Clark, A. and Clark, D. and Compostella, G. and Convery, M. E. and Conway, J. and Cooper, B. and Copic, K. and Cordelli, M. and Cortiana, G. and Crescioli, F. and Cuenca Almenar, C. and Cuevas, J. and Culbertson, R. and Cully, J. C. and Dagenhart, D. and Datta, M. and Davies, T. and de Barbaro, P. and De Cecco, S. and Deisher, A. and De Lentdecker, G. and De Lorenzo, G. and Dell'Orso, M. and Demortier, L. and Deng, J. and Deninno, M. and De Pedis, D. and Derwent, P. F. and Di Giovanni, G. P. and Dionisi, C. and Di Ruzza, B. and Dittmann, J. R. and D'Onofrio, M. and Donati, S. and Dong, P. and Donini, J. and Dorigo, T. and Dube, S. and Efron, J. and Erbacher, R. and Errede, D. and Errede, S. and Eusebi, R. and Fang, H. C. and Farrington, S. and Fedorko, W. T. and Feild, R. G. and Feindt, M. and Fernandez, J. P. and Ferrazza, C. and Field, R. and Flanagan, G. and Forrest, R. and Forrester, S. and Franklin, M. and Freeman, J. C. and Furic, I. and Gallinaro, M. and Galyardt, J. and Garberson, F. and Garcia, J. E. and Garfinkel, A. F. and Genser, K. and Gerberich, H. and Gerdes, D. and Giagu, S. and Giakoumopolou, V. and Giannetti, P. and Gibson, K. and Gimmell, J. L. and Ginsburg, C. M. and Giokaris, N. and Giordani, M. and Giromini, P. and Giunta, M. and Glagolev, V. and Glenzinski, D. and Gold, M. and Goldschmidt, N. and Golossanov, A. and Gomez, G. and Gomez-Ceballos, G. and Goncharov, M. and Gonz\'alez, O. and Gorelov, I. and Goshaw, A. T. and Goulianos, K. and Gresele, A. and Grinstein, S. and Grosso-Pilcher, C. and Group, R. C. and Grundler, U. and Guimaraes da Costa, J. and Gunay-Unalan, Z. and Haber, C. and Hahn, K. and Hahn, S. R. and Halkiadakis, E. and Hamilton, A. and Han, B.-Y. and Han, J. Y. and Handler, R. and Happacher, F. and Hara, K. and Hare, D. and Hare, M. and Harper, S. and Harr, R. F. and Harris, R. M. and Hartz, M. and Hatakeyama, K. and Hauser, J. and Hays, C. and Heck, M. and Heijboer, A. and Heinemann, B. and Heinrich, J. and Henderson, C. and Herndon, M. and Heuser, J. and Hewamanage, S. and Hidas, D. and Hill, C. S. and Hirschbuehl, D. and Hocker, A. and Hou, S. and Houlden, M. and Hsu, S.-C. and Huffman, B. T. and Hughes, R. E. and Husemann, U. and Huston, J. and Incandela, J. and Introzzi, G. and Iori, M. and Ivanov, A. and Iyutin, B. and James, E. and Jayatilaka, B. and Jeans, D. and Jeon, E. J. and Jindariani, S. and Johnson, W. and Jones, M. and Joo, K. K. and Jun, S. Y. and Jung, J. E. and Junk, T. R. and Kamon, T. and Kar, D. and Karchin, P. E. and Kato, Y. and Kephart, R. and Kerzel, U. and Khotilovich, V. and Kilminster, B. and Kim, D. H. and Kim, H. S. and Kim, J. E. and Kim, M. J. and Kim, S. B. and Kim, S. H. and Kim, Y. K. and Kimura, N. and Kirsch, L. and Klimenko, S. and Klute, M. and Knuteson, B. and Ko, B. R. and Koay, S. A. and Kondo, K. and Kong, D. J. and Konigsberg, J. and Korytov, A. and Kotwal, A. V. and Kraus, J. and Kreps, M. and Kroll, J. and Krumnack, N. and Kruse, M. and Krutelyov, V. and Kubo, T. and Kuhlmann, S. E. and Kuhr, T. and Kulkarni, N. P. and Kusakabe, Y. and Kwang, S. and Laasanen, A. T. and Lai, S. and Lami, S. and Lammel, S. and Lancaster, M. and Lander, R. L. and Lannon, K. and Lath, A. and Latino, G. and Lazzizzera, I. and LeCompte, T. and Lee, J. and Lee, J. and Lee, Y. J. and Lee, S. W. and Lef\`evre, R. and Leonardo, N. and Leone, S. and Levy, S. and Lewis, J. D. and Lin, C. and Lin, C. S. and Linacre, J. and Lindgren, M. and Lipeles, E. and Lister, A. and Litvintsev, D. O. and Liu, T. and Lockyer, N. S. and Loginov, A. and Loreti, M. and Lovas, L. and Lu, R.-S. and Lucchesi, D. and Lueck, J. and Luci, C. and Lujan, P. and Lukens, P. and Lungu, G. and Lyons, L. and Lys, J. and Lysak, R. and Lytken, E. and Mack, P. and MacQueen, D. and Madrak, R. and Maeshima, K. and Makhoul, K. and Maki, T. and Maksimovic, P. and Malde, S. and Malik, S. and Manca, G. and Manousakis, A. and Margaroli, F. and Marino, C. and Marino, C. P. and Martin, A. and Martin, M. and Martin, V. and Mart\'{\i}nez, M. and Mart\'{\i}nez-Ballar\'{\i}n, R. and Maruyama, T. and Mastrandrea, P. and Masubuchi, T. and Mattson, M. E. and Mazzanti, P. and McFarland, K. S. and McIntyre, P. and McNulty, R. and Mehta, A. and Mehtala, P. and Menzemer, S. and Menzione, A. and Merkel, P. and Mesropian, C. and Messina, A. and Miao, T. and Miladinovic, N. and Miles, J. and Miller, R. and Mills, C. and Milnik, M. and Mitra, A. and Mitselmakher, G. and Miyake, H. and Moed, S. and Moggi, N. and Moon, C. S. and Moore, R. and Morello, M. and Movilla Fernandez, P. and M\"ulmenst\"adt, J. and Mukherjee, A. and Muller, Th. and Mumford, R. and Murat, P. and Mussini, M. and Nachtman, J. and Nagai, Y. and Nagano, A. and Naganoma, J. and Nakamura, K. and Nakano, I. and Napier, A. and Necula, V. and Neu, C. and Neubauer, M. S. and Nielsen, J. and Nodulman, L. and Norman, M. and Norniella, O. and Nurse, E. and Oh, S. H. and Oh, Y. D. and Oksuzian, I. and Okusawa, T. and Oldeman, R. and Orava, R. and Osterberg, K. and Pagan Griso, S. and Pagliarone, C. and Palencia, E. and Papadimitriou, V. and Papaikonomou, A. and Paramonov, A. A. and Parks, B. and Pashapour, S. and Patrick, J. and Pauletta, G. and Paulini, M. and Paus, C. and Pellett, D. E. and Penzo, A. and Phillips, T. J. and Piacentino, G. and Piedra, J. and Pinera, L. and Pitts, K. and Plager, C. and Pondrom, L. and Portell, X. and Poukhov, O. and Pounder, N. and Prakoshyn, F. and Pronko, A. and Proudfoot, J. and Ptohos, F. and Punzi, G. and Pursley, J. and Rademacker, J. and Rahaman, A. and Ramakrishnan, V. and Ranjan, N. and Redondo, I. and Reisert, B. and Rekovic, V. and Renton, P. and Rescigno, M. and Richter, S. and Rimondi, F. and Ristori, L. and Robson, A. and Rodrigo, T. and Rogers, E. and Rolli, S. and Roser, R. and Rossi, M. and Rossin, R. and Roy, P. and Ruiz, A. and Russ, J. and Rusu, V. and Saarikko, H. and Safonov, A. and Sakumoto, W. K. and Salamanna, G. and Salt\'o, O. and Santi, L. and Sarkar, S. and Sartori, L. and Sato, K. and Savoy-Navarro, A. and Scheidle, T. and Schlabach, P. and Schmidt, E. E. and Schmidt, M. A. and Schmidt, M. P. and Schmitt, M. and Schwarz, T. and Scodellaro, L. and Scott, A. L. and Scribano, A. and Scuri, F. and Sedov, A. and Seidel, S. and Seiya, Y. and Semenov, A. and Sexton-Kennedy, L. and Sfyria, A. and Shalhout, S. Z. and Shapiro, M. D. and Shears, T. and Shepard, P. F. and Sherman, D. and Shimojima, M. and Shochet, M. and Shon, Y. and Shreyber, I. and Sidoti, A. and Sinervo, P. and Sisakyan, A. and Slaughter, A. J. and Slaunwhite, J. and Sliwa, K. and Smith, J. R. and Snider, F. D. and Snihur, R. and Soderberg, M. and Soha, A. and Somalwar, S. and Sorin, V. and Spalding, J. and Spinella, F. and Spreitzer, T. and Squillacioti, P. and Stanitzki, M. and St. Denis, R. and Stelzer, B. and Stelzer-Chilton, O. and Stentz, D. and Strologas, J. and Stuart, D. and Suh, J. S. and Sukhanov, A. and Sun, H. and Suslov, I. and Suzuki, T. and Taffard, A. and Takashima, R. and Takeuchi, Y. and Tanaka, R. and Tecchio, M. and Teng, P. K. and Terashi, K. and Thom, J. and Thompson, A. S. and Thompson, G. A. and Thomson, E. and Tipton, P. and Tiwari, V. and Tkaczyk, S. and Toback, D. and Tokar, S. and Tollefson, K. and Tomura, T. and Tonelli, D. and Torre, S. and Torretta, D. and Tourneur, S. and Trischuk, W. and Tu, Y. and Turini, N. and Ukegawa, F. and Uozumi, S. and Vallecorsa, S. and van Remortel, N. and Varganov, A. and Vataga, E. and V\'azquez, F. and Velev, G. and Vellidis, C. and Veszpremi, V. and Vidal, M. and Vidal, R. and Vila, I. and Vilar, R. and Vine, T. and Vogel, M. and Volobouev, I. and Volpi, G. and W\"urthwein, F. and Wagner, P. and Wagner, R. G. and Wagner, R. L. and Wagner-Kuhr, J. and Wagner, W. and Wakisaka, T. and Wallny, R. and Wang, S. M. and Warburton, A. and Waters, D. and Weinberger, M. and Wester, W. C. and Whitehouse, B. and Whiteson, D. and Wicklund, A. B. and Wicklund, E. and Williams, G. and Williams, H. H. and Wilson, P. and Winer, B. L. and Wittich, P. and Wolbers, S. and Wolfe, C. and Wright, T. and Wu, X. and Wynne, S. M. and Yagil, A. and Yamamoto, K. and Yamaoka, J. and Yamashita, T. and Yang, C. and Yang, U. K. and Yang, Y. C. and Yao, W. M. and Yeh, G. P. and Yoh, J. and Yorita, K. and Yoshida, T. and Yu, G. B. and Yu, I. and Yu, S. S. and Yun, J. C. and Zanello, L. and Zanetti, A. and Zaw, I. and Zhang, X. and Zheng, Y. and Zucchelli, S.},
  collaboration = {CDF Collaboration},
  journal = "\href{https://link.aps.org/doi/10.1103/PhysRevLett.100.182002}{Phys. Rev. Lett.}",
  volume = "{100}",
  issue = "{18}",
  pages = "{182002}",
  numpages = "{7}",
  year = "{2008}",
  month = {May},
  publisher = {American Physical Society},
  doi = {10.1103/PhysRevLett.100.182002},
  url = {https://link.aps.org/doi/10.1103/PhysRevLett.100.182002}
}

@article{abazov2008,
  title = {Observation of the ${B}_{c}$ Meson in the Exclusive Decay ${B}_{c}\ensuremath{\rightarrow}J/\ensuremath{\psi}\ensuremath{\pi}$},
  author = {Abazov, V. M. and Abbott, B. and Abolins, M. and Acharya, B. S. and Adams, M. and Adams, T. and Aguilo, E. and Ahn, S. H. and Ahsan, M. and Alexeev, G. D. and Alkhazov, G. and Alton, A. and Alverson, G. and Alves, G. A. and Anastasoaie, M. and Ancu, L. S. and Andeen, T. and Anderson, S. and Andrieu, B. and Anzelc, M. S. and Aoki, M. and Arnoud, Y. and Arov, M. and Arthaud, M. and Askew, A. and \AA{}sman, B. and Assis Jesus, A. C. S. and Atramentov, O. and Avila, C. and Ay, C. and Badaud, F. and Baden, A. and Bagby, L. and Baldin, B. and Bandurin, D. V. and Banerjee, P. and Banerjee, S. and Barberis, E. and Barfuss, A.-F. and Bargassa, P. and Baringer, P. and Barreto, J. and Bartlett, J. F. and Bassler, U. and Bauer, D. and Beale, S. and Bean, A. and Begalli, M. and Begel, M. and Belanger-Champagne, C. and Bellantoni, L. and Bellavance, A. and Benitez, J. A. and Beri, S. B. and Bernardi, G. and Bernhard, R. and Bertram, I. and Besan\ifmmode \mbox{\c{c}}\else \c{c}\fi{}on, M. and Beuselinck, R. and Bezzubov, V. A. and Bhat, P. C. and Bhatnagar, V. and Biscarat, C. and Blazey, G. and Blekman, F. and Blessing, S. and Bloch, D. and Bloom, K. and Boehnlein, A. and Boline, D. and Bolton, T. A. and Borissov, G. and Bose, T. and Brandt, A. and Brock, R. and Brooijmans, G. and Bross, A. and Brown, D. and Buchanan, N. J. and Buchholz, D. and Buehler, M. and Buescher, V. and Bunichev, V. and Burdin, S. and Burke, S. and Burnett, T. H. and Buszello, C. P. and Butler, J. M. and Calfayan, P. and Calvet, S. and Cammin, J. and Carvalho, W. and Casey, B. C. K. and Castilla-Valdez, H. and Chakrabarti, S. and Chakraborty, D. and Chan, K. and Chan, K. M. and Chandra, A. and Charles, F. and Cheu, E. and Chevallier, F. and Cho, D. K. and Choi, S. and Choudhary, B. and Christofek, L. and Christoudias, T. and Cihangir, S. and Claes, D. and Coadou, Y. and Cooke, M. and Cooper, W. E. and Corcoran, M. and Couderc, F. and Cousinou, M.-C. and Cr\'ep\'e-Renaudin, S. and Cutts, D. and \ifmmode \acute{C}\else \'{C}\fi{}wiok, M. and da Motta, H. and Das, A. and Davies, G. and De, K. and de Jong, S. J. and De La Cruz-Burelo, E. and De Oliveira Martins, C. and Degenhardt, J. D. and D\'eliot, F. and Demarteau, M. and Demina, R. and Denisov, D. and Denisov, S. P. and Desai, S. and Diehl, H. T. and Diesburg, M. and Dominguez, A. and Dong, H. and Dudko, L. V. and Duflot, L. and Dugad, S. R. and Duggan, D. and Duperrin, A. and Dyer, J. and Dyshkant, A. and Eads, M. and Edmunds, D. and Ellison, J. and Elvira, V. D. and Enari, Y. and Eno, S. and Ermolov, P. and Evans, H. and Evdokimov, A. and Evdokimov, V. N. and Ferapontov, A. V. and Ferbel, T. and Fiedler, F. and Filthaut, F. and Fisher, W. and Fisk, H. E. and Fortner, M. and Fox, H. and Fu, S. and Fuess, S. and Gadfort, T. and Galea, C. F. and Gallas, E. and Garcia, C. and Garcia-Bellido, A. and Gavrilov, V. and Gay, P. and Geist, W. and Gel\'e, D. and Gerber, C. E. and Gershtein, Y. and Gillberg, D. and Ginther, G. and Gollub, N. and G\'omez, B. and Goussiou, A. and Grannis, P. D. and Greenlee, H. and Greenwood, Z. D. and Gregores, E. M. and Grenier, G. and Gris, Ph. and Grivaz, J.-F. and Grohsjean, A. and Gr\"unendahl, S. and Gr\"unewald, M. W. and Guo, F. and Guo, J. and Gutierrez, G. and Gutierrez, P. and Haas, A. and Hadley, N. J. and Haefner, P. and Hagopian, S. and Haley, J. and Hall, I. and Hall, R. E. and Han, L. and Harder, K. and Harel, A. and Harrington, R. and Hauptman, J. M. and Hauser, R. and Hays, J. and Hebbeker, T. and Hedin, D. and Hegeman, J. G. and Heinmiller, J. M. and Heinson, A. P. and Heintz, U. and Hensel, C. and Herner, K. and Hesketh, G. and Hildreth, M. D. and Hirosky, R. and Hobbs, J. D. and Hoeneisen, B. and Hoeth, H. and Hohlfeld, M. and Hong, S. J. and Hossain, S. and Houben, P. and Hu, Y. and Hubacek, Z. and Hynek, V. and Iashvili, I. and Illingworth, R. and Ito, A. S. and Jabeen, S. and Jaffr\'e, M. and Jain, S. and Jakobs, K. and Jarvis, C. and Jesik, R. and Johns, K. and Johnson, C. and Johnson, M. and Jonckheere, A. and Jonsson, P. and Juste, A. and Kajfasz, E. and Kalinin, A. M. and Kalk, J. M. and Kappler, S. and Karmanov, D. and Kasper, P. A. and Katsanos, I. and Kau, D. and Kaushik, V. and Kehoe, R. and Kermiche, S. and Khalatyan, N. and Khanov, A. and Kharchilava, A. and Kharzheev, Y. M. and Khatidze, D. and Kim, T. J. and Kirby, M. H. and Kirsch, M. and Klima, B. and Kohli, J. M. and Konrath, J.-P. and Korablev, V. M. and Kozelov, A. V. and Kraus, J. and Krop, D. and Kuhl, T. and Kumar, A. and Kupco, A. and Kur\ifmmode \check{c}\else \v{c}\fi{}a, T. and Kvita, J. and Lacroix, F. and Lam, D. and Lammers, S. and Landsberg, G. and Lebrun, P. and Lee, W. M. and Leflat, A. and Lellouch, J. and Leveque, J. and Li, J. and Li, L. and Li, Q. Z. and Lietti, S. M. and Lima, J. G. R. and Lincoln, D. and Linnemann, J. and Lipaev, V. V. and Lipton, R. and Liu, Y. and Liu, Z. and Lobodenko, A. and Lokajicek, M. and Love, P. and Lubatti, H. J. and Luna, R. and Lyon, A. L. and Maciel, A. K. A. and Mackin, D. and Madaras, R. J. and M\"attig, P. and Magass, C. and Magerkurth, A. and Mal, P. K. and Malbouisson, H. B. and Malik, S. and Malyshev, V. L. and Mao, H. S. and Maravin, Y. and Martin, B. and McCarthy, R. and Melnitchouk, A. and Mendoza, L. and Mercadante, P. G. and Merkin, M. and Merritt, K. W. and Meyer, A. and Meyer, J. and Millet, T. and Mitrevski, J. and Molina, J. and Mommsen, R. K. and Mondal, N. K. and Moore, R. W. and Moulik, T. and Muanza, G. S. and Mulders, M. and Mulhearn, M. and Mundal, O. and Mundim, L. and Nagy, E. and Naimuddin, M. and Narain, M. and Naumann, N. A. and Neal, H. A. and Negret, J. P. and Neustroev, P. and Nilsen, H. and Nogima, H. and Novaes, S. F. and Nunnemann, T. and O'Dell, V. and O'Neil, D. C. and Obrant, G. and Ochando, C. and Onoprienko, D. and Oshima, N. and Osman, N. and Osta, J. and Otec, R. and Otero y Garz\'on, G. J. and Owen, M. and Padley, P. and Pangilinan, M. and Parashar, N. and Park, S.-J. and Park, S. K. and Parsons, J. and Partridge, R. and Parua, N. and Patwa, A. and Pawloski, G. and Penning, B. and Perfilov, M. and Peters, K. and Peters, Y. and P\'etroff, P. and Petteni, M. and Piegaia, R. and Piper, J. and Pleier, M.-A. and Podesta-Lerma, P. L. M. and Podstavkov, V. M. and Pogorelov, Y. and Pol, M.-E. and Polozov, P. and Pope, B. G. and Popov, A. V. and Potter, C. and Prado da Silva, W. L. and Prosper, H. B. and Protopopescu, S. and Qian, J. and Quadt, A. and Quinn, B. and Rakitine, A. and Rangel, M. S. and Ranjan, K. and Ratoff, P. N. and Renkel, P. and Reucroft, S. and Rich, P. and Rieger, J. and Rijssenbeek, M. and Ripp-Baudot, I. and Rizatdinova, F. and Robinson, S. and Rodrigues, R. F. and Rominsky, M. and Royon, C. and Rubinov, P. and Ruchti, R. and Safronov, G. and Sajot, G. and S\'anchez-Hern\'andez, A. and Sanders, M. P. and Santoro, A. and Savage, G. and Sawyer, L. and Scanlon, T. and Schaile, D. and Schamberger, R. D. and Scheglov, Y. and Schellman, H. and Schliephake, T. and Schwanenberger, C. and Schwartzman, A. and Schwienhorst, R. and Sekaric, J. and Severini, H. and Shabalina, E. and Shamim, M. and Shary, V. and Shchukin, A. A. and Shivpuri, R. K. and Siccardi, V. and Simak, V. and Sirotenko, V. and Skubic, P. and Slattery, P. and Smirnov, D. and Snow, G. R. and Snow, J. and Snyder, S. and S\"oldner-Rembold, S. and Sonnenschein, L. and Sopczak, A. and Sosebee, M. and Soustruznik, K. and Spurlock, B. and Stark, J. and Steele, J. and Stolin, V. and Stoyanova, D. A. and Strandberg, J. and Strandberg, S. and Strang, M. A. and Strauss, E. and Strauss, M. and Str\"ohmer, R. and Strom, D. and Stutte, L. and Sumowidagdo, S. and Svoisky, P. and Sznajder, A. and Tamburello, P. and Tanasijczuk, A. and Taylor, W. and Temple, J. and Tiller, B. and Tissandier, F. and Titov, M. and Tokmenin, V. V. and Toole, T. and Torchiani, I. and Trefzger, T. and Tsybychev, D. and Tuchming, B. and Tully, C. and Tuts, P. M. and Unalan, R. and Uvarov, L. and Uvarov, S. and Uzunyan, S. and Vachon, B. and van den Berg, P. J. and Van Kooten, R. and van Leeuwen, W. M. and Varelas, N. and Varnes, E. W. and Vasilyev, I. A. and Vaupel, M. and Verdier, P. and Vertogradov, L. S. and Verzocchi, M. and Villeneuve-Seguier, F. and Vint, P. and Vokac, P. and Von Toerne, E. and Voutilainen, M. and Wagner, R. and Wahl, H. D. and Wang, L. and Wang, M. H. L. S. and Warchol, J. and Watts, G. and Wayne, M. and Weber, G. and Weber, M. and Welty-Rieger, L. and Wenger, A. and Wermes, N. and Wetstein, M. and White, A. and Wicke, D. and Wilson, G. W. and Wimpenny, S. J. and Wobisch, M. and Wood, D. R. and Wyatt, T. R. and Xie, Y. and Yacoob, S. and Yamada, R. and Yan, M. and Yasuda, T. and Yatsunenko, Y. A. and Yip, K. and Yoo, H. D. and Youn, S. W. and Yu, J. and Zatserklyaniy, A. and Zeitnitz, C. and Zhao, T. and Zhou, B. and Zhu, J. and Zielinski, M. and Zieminska, D. and Zieminski, A. and Zivkovic, L. and Zutshi, V. and Zverev, E. G.},
  collaboration = {D0 Collaboration},
  journal = "\href{https://link.aps.org/doi/10.1103/PhysRevLett.101.012001}{Phys. Rev. Lett.}",
  volume = "{101}",
  issue = "{1}",
  pages = "{012001}",
  numpages = "{6}",
  year = "{2008}",
  month = {Jul},
  publisher = {American Physical Society},
  doi = {10.1103/PhysRevLett.101.012001},
  url = {https://link.aps.org/doi/10.1103/PhysRevLett.101.012001}
}

@article{godfrey2016b,
    author = "Godfrey, Stephen and Moats, K. and Swanson, E. S.",
    title = "{$B$ and $B_s$ Meson Spectroscopy}",
    eprint = "1607.02169",
    archivePrefix = "arXiv",
    primaryClass = "hep-ph",
    doi = "10.1103/PhysRevD.94.054025",
    journal = "Phys. Rev. D",
    volume = "94",
    number = "5",
    pages = "054025",
    year = "2016"
}

@article{aaij2015b,
    author = "Aaij, Roel and others",
    collaboration = "LHCb",
    title = "{Precise measurements of the properties of the $B_1(5721)^{0,+}$ and $B^\ast_2(5747)^{0,+}$ states and observation of $B^{+,0}\pi^{-,+}$ mass structures}",
    eprint = "1502.02638",
    archivePrefix = "arXiv",
    primaryClass = "hep-ex",
    reportNumber = "LHCB-PAPER-2014-067, CERN-PH-EP-2015-021",
    doi = "10.1007/JHEP04(2015)024",
    journal = "JHEP",
    volume = "04",
    pages = "024",
    year = "2015"
}

@article{aaltonen2009,
    author = "Aaltonen, T. and others",
    collaboration = "CDF",
    title = "{Measurement of Resonance Parameters of Orbitally Excited Narrow $B^0$ Mesons}",
    eprint = "0809.5007",
    archivePrefix = "arXiv",
    primaryClass = "hep-ex",
    reportNumber = "FERMILAB-PUB-08-416-E",
    doi = "10.1103/PhysRevLett.102.102003",
    journal = "Phys. Rev. Lett.",
    volume = "102",
    pages = "102003",
    year = "2009"
}

@article{aaij2013,
    author = "Aaij, R and others",
    collaboration = "LHCb",
    title = "{First observation of the decay $B_{s2}^*(5840)^0 \to B^{*+} K^-$ and studies of excited $B^0_s$ mesons}",
    eprint = "1211.5994",
    archivePrefix = "arXiv",
    primaryClass = "hep-ex",
    reportNumber = "LHCB-PAPER-2012-030, CERN-PH-EP-2012-340",
    doi = "10.1103/PhysRevLett.110.151803",
    journal = "Phys. Rev. Lett.",
    volume = "110",
    number = "15",
    pages = "151803",
    year = "2013"
}

@article{aaltonen2014,
    author = "Aaltonen, Timo Antero and others",
    collaboration = "CDF",
    title = "{Study of Orbitally Excited $B$ Mesons and Evidence for a New $B\pi$ Resonance}",
    eprint = "1309.5961",
    archivePrefix = "arXiv",
    primaryClass = "hep-ex",
    reportNumber = "FERMILAB-PUB-13-393-E, PUB-13-393-E-(FERMILAB)",
    doi = "10.1103/PhysRevD.90.012013",
    journal = "Phys. Rev. D",
    volume = "90",
    number = "1",
    pages = "012013",
    year = "2014"
}
\bibliographystyle{epj}
\end{document}